\documentclass{aa}
\usepackage{txfonts}
\usepackage{graphicx}
\usepackage{subfigure}

\def\apj{ApJ}
\def\apjl{ApJ}
\def\apjs{ApJS}
\def\ao{Appl.~Opt.}

\def\aap{A\&A}

\def\mnras{MNRAS}
\def\Ms{\mbox{${\rm M}_{\odot}$}}

\def\Msyr{\mbox{\Ms yr$^{-1}$}}
\def\Msyr{\mbox{\Ms yr$^{-1}$}}

\def\lya{\mbox{${\rm Ly}{\alpha}$}}
\def\ewlya{\mbox{$EW{\rm{(Ly}\alpha}\rm{)}$}}
\def\ewlyaem{\mbox{$EW(\rm{Ly}\alpha_{\rm{EM}})$}}

\def\sfrlya{\mbox{$SFR_{\rm{Ly}\alpha}$}}
\def\sfruv{\mbox{$SFR_{\rm{UV}}$}}

\def\flya{\mbox{$F_{\rm{Ly}\alpha}$}}

\def\twwattm{\mbox{10$^{-21}$Wm$^{-2} $}}

\def\ll{\mbox{${\lambda \lambda}$}}

\def\Zs{\mbox{${\rm Z}_{\odot}$}}
\def\mR{\mbox{$m_{\rm{R}}$}}
\def\mg{\mbox{$m_{\rm{g}}$}}
\def\kms{\mbox{$\rm{km} s^{-1}$}}

\def\nhi{\mbox{$N_{\rm{HI}}$}}

\def\zphot{\mbox{$z_{\rm{phot} }$}}

\def\ewlis{\mbox{$EW\rm{(LIS)}$}}
\def\fwhmlis{\mbox{$FWHM\rm{(LIS)}$}}

\begin{document}
\title{\lya\ emission in high-redshift galaxies \thanks{Based on observations
    (proposals: 069.A-0105(A) and 071.A-0307(A)) obtained at the ESO VLT at
    Cerro Paranal, Chile, and on observations made with HST ACS (GO proposal:
    9502).}}  \subtitle{} \author{ C.~Tapken\inst{1,2} \and
  I.~Appenzeller\inst{1} \and S.~Noll\inst{3} \and S.~Richling\inst{4} \and
  J.~Heidt\inst{1} \and E.~Meink\"ohn\inst{5} \and D.~Mehlert\inst{2} }
\institute{Landessternwarte Heidelberg-K\"onigstuhl, D-69117 Heidelberg,
  Germany
%\and Universit\"ats-Sternwarte M\"unchen, Scheinerstr. 1, D-81679, M\"unchen, Germany
  \and Max-Planck-Institut f\"ur Astronomie, K\"onigstuhl 17, D-69117
  Heidelberg, Germany \and Max-Planck-Institut f\"ur extraterrestrische
  Physik, Giessenbachstr., D-85741 Garching, Germany \and Institut d'
  Astrophysique de Paris, 98bis Bd Arago, 75014 Paris, France \and Institut
  f\"ur Theoretische Astrophysik, Albert-Ueberle-Strasse 2, D-69120
  Heidelberg, Germany }

\offprints{C. Tapken, Heidelberg (\email{tapken@mpia.de})} \date{received;
  accepted}

\abstract{ A significant fraction of the high-redshift galaxies show strong
  \lya\ emission lines. For redshifts $z>5$, most known galaxies belong to this
  class. However, so far not much is known about the physical structure and
  nature of these objects.}{Our aim is to analyse the \lya\ emission in a
  sample of high-redshift UV-continuum selected galaxies and to derive the
  physical conditions that determine the \lya\ profile and the line
  strength.}{ VLT/FORS spectra with a resolution of $R \approx$ 2000 of 16
  galaxies in the redshift range of $z = 2.7$ to $5$ are presented. The
  observed \lya\ profiles are compared with theoretical models.} {The \lya\ 
  lines range from pure absorption ($EW$ = -17 \AA ) to strong emission ($EW$
  = 153 \AA ).  Most \lya\ emission lines show an asymmetric profile, and three
  galaxies have a double-peaked profile.  Both types of profiles can be
  explained by a uniform model consisting of an expanding shell of neutral and
  ionised hydrogen around a compact starburst region.  The broad, blueshifted,
  low-ionisation interstellar absorption lines indicate a galaxy-scale outflow
  of the ISM. The strengths of these lines are found to be determined in part 
  by the velocity dispersion of the outflowing medium.  We find star-formation
  rates of these galaxies ranging from $SFR_{\rm{UV}}$ = 1.2 to 63.2 \Msyr\ 
  uncorrected for dust absorption.}  {The \lya\ emission strength of our
  target galaxies is found to be determined by the amount of dust and the
  kinematics of the outflowing material.}


\maketitle \titlerunning{\lya\ emission in high-redshift galaxies }
\authorrunning{C.~Tapken et al.}
\section{Introduction}
The 8-10 meter class telescopes and the Hubble Space Telescope have made it
feasible to investigate details of the galaxies in the young universe. The
majority of the high-redshift galaxies were detected on the basis of their UV
colours.  Notable examples are the so-called Lyman-break galaxies, which are
selected using continuum breaks between 912 and 1216 \AA\ (for a review see
Giavalisco \cite{giavalisco2002}). Since a selection by the Lyman break proved
to be very efficient in terms of telescope time, a large sample of
high-redshift galaxies has been derived. Shapley et al. (\cite{shapley2003}),
e.g., presented the spectra of 798 galaxies at a redshift of $\approx$ 3.
Alternatively, many galaxies, especially in the $z>5$ universe, have been found
by means of their strong \lya\ emission (e.g., Hu et al. \cite{hu2004}), the
so-called ``\lya\ emission line galaxies''(LAEs).

Well-defined gaps in the telluric OH-bands allow one to detect LAEs rather
efficiently from their excess in narrow-band filters (e.g., Hu et al.
\cite{hu1998}; Kudritzki et al. \cite{kudritzki2000}; Rhoads et al. \cite{rhoads2000}; Maier et al. \cite{maier2003};
Ouchi et al. \cite{ouchi2005}; Tapken et al. \cite{tapken2006}; Nilsson et
al. \cite{nilsson2006}). The frequency
and properties of the LAEs have been used to derive the luminosity function
(Hu et al. \cite{hu2004}), the star-formation rate in the early universe
(Ajiki et al. \cite{ajiki2003}), the epoch of re-ionisation (Rhoads et al.
\cite{rhoads2004}), and led to the discovery of a large-scale structure at
z$\approx$5.7 (Ouchi et al. \cite{ouchi2005}).  Moreover, their luminosity
function can be compared to theoretical models (Haiman \& Spaans
\cite{haiman1999}; Thommes \& Meisenheimer \cite{thommes2005}; Le Delliou et
al. \cite{ledelliou2005}), allowing models of galaxy formation
and evolution to be constrained.  LAEs are the most distant objects in the
universe, to have been detected so far (Iye et al. \cite{iye2006}).

Star-forming galaxies produce huge amounts of \lya\ photons by recombination
in HII regions that are ionised by young stars (e.g., Charlot \& Fall
\cite{charlot1993}).  However, the fact that \lya\ is a resonance line makes
physical interpretation of the \lya\ emission challenging.  Many theoretical
efforts have been made to treat the radiation transfer of resonance lines,
both numerically (e.g., Auer \cite{auer1968}; Adams \cite{adams1972}; Ahn et
al. \cite{ahn2001}; Richling \cite{richling2003}; Hansen \& Oh
\cite{hansen2006}; Verhamme et al. \cite{verhamme2006}) and analytically
(e.g., Osterbrock \cite{osterbrock1962}; Neufeld \cite{neufeld1990}).  Neutral
hydrogen absorbs and re-emits \lya\ photons in random directions. As a result the
\lya\ photons will diffuse in spatial and frequency space (see, e.g., Neufeld
\cite{neufeld1990}). The diffusion in frequency space changes the intrinsic
\lya\ profile, leading to complex \lya\ profile morphologies. The diffusion in
space increases the optical path of the \lya\ compared to UV-continuum
photons. Therefore, any dust distributed uniformly in a neutral medium affects the
\lya\ photons more strongly than the UV-continuum photons. This could explain
the fact that most UV-continuum selected high-redshift galaxies show
weak \lya\ emission or none at all (Shapley et al. \cite{shapley2003}; Noll et al.
\cite{noll2004}).  As demonstrated by Neufeld (\cite{neufeld1991}) and Hansen
\& Oh (\cite{hansen2006}), the spatial distribution of the dust relative to the
neutral hydrogen also plays an important role as the \lya\ emission can even
be less affected by dust than the continuum radiation in a two-phase
interstellar medium. Moreover, a large-scale outflow of neutral gas can
decrease the number of \lya\ resonance scattering (Kunth et al.
\cite{kunth1998}), if the neutral gas is velocity-shifted with respect to the
ionised gas. To conclude, interpretation of the flux, profile, and spatial
morphology of the \lya\ line is not straightforward.  Without further insight
into the \lya\ emission of star-forming galaxies, the use of LAEs as a tool
for constraining models of galaxy formation is limited.

Information about the formation and nature of the \lya\ emission in
high-redshift galaxies can be derived from the analysis of the UV-restframe
continuum, as demonstrated by Shapley et al. (\cite{shapley2003}).  Using
average spectra (the so-called composite spectra) of UV-continuum selected
high-redshift galaxies, Shapley et al. (\cite{shapley2003}) and Noll et al.
(\cite{noll2004}) could show that the strength of the \lya\ emission
correlates well with other galaxy properties, such as the slope of the
continuum or the strength of the interstellar absorption lines. These
correlations may be explained by varying amounts of dust and by the kinematics
of the host galaxies.  Another method for deriving information about the nature
of the \lya\ emission is to analyse the \lya\ profile. Dawson et al.
(\cite{dawson2002}) and Westra et al. (\cite{westra2005}) demonstrated that
the profile of the \lya\ emission can be used to derive properties of the
emitting galaxy. The detailed comparison of \lya\ profiles with dedicated
radiative transfer models can constrain the kinematics of the emitting and
neutral ISM. However, the spectra of Shapley et al. (\cite{shapley2003}) and
Noll et al. (\cite{noll2004}) had too low a spectral resolution to derive
kinematical information from the \lya\ profile.

In this paper we analyse a sample of high-redshift UV-continuum selected
galaxies in the FORS Deep Field (FDF) to constrain the origin of the \lya\ 
emission, the evolutionary state, and the physical properties of these
galaxies.  The FORS Deep Field (FDF) is a deep photometric and spectroscopic
survey (Appenzeller et al. \cite{appenzeller2000}; Heidt et al.
\cite{heidt2003}; Noll et al. \cite{noll2004}) carried out with the FORS
instruments at the ESO Very Large Telescope (VLT).  The analysis of the galaxy
sample of this work has made use of the FDF photometric and spectroscopic
surveys and the HST-imaging follow-up.  In addition R $\approx$ 2000 spectra
were obtained, which include the \lya\ line.  The data are presented in Sect.
\ref{sec_data}. The properties of the galaxies, including the strength and
profile of the \lya\ lines, the star-formation rates, and the interstellar
absorption lines, are presented in Sect. \ref{sec_results}. The \lya\ profiles
are compared to theoretical models in Sect. \ref{sec_modellya}. The results
are discussed in Sect. \ref{sec_discussion} and the conclusions are given in
Sect. \ref{sec_conclusion}.

Throughout this paper we adopt $\Omega_{\Lambda}$ = 0.7, $\Omega_M$ = 0.3, and
H$_{0}$ = 70 km s$^{-1}$Mpc$^{-1}$. The magnitudes are given in the Vega
system.

\section{Presentation of the data}\label{sec_data} 

\subsection{Photometric data}\label{sec:phot_data} 
The FDF survey consists of deep optical U, B, g, R, I observations obtained
with FORS at the VLT and near-infrared J and Ks observations obtained with
SofI at the ESO-NTT (Heidt et al. \cite{heidt2003}). In addition, deep
observations with an SDSS z and a medium-band filter centred at 8350 \AA\ were
obtained (Gabasch et al. \cite{gabasch2004}). The FDF was imaged in the
broad-band F814W filter with the Advanced Camera for Surveys (ACS) on the
Hubble Space Telescope (HST) in autumn 2002. The field was covered with 4 ACS
pointings of 40 minute exposure each, reaching a 10$\sigma$ limit of 25.6
mag.
%AB > vega 26.0 > 25.6 
The data reduction was performed with the standard {\sl CALACS\footnote
  {www.stsci.edu/hst/acs/analysis}} pipeline, and the combined final mosaic
was produced with the multi-drizzle package (Mutchler et al.
\cite{mutchler2003}). The final combined image has a pixel scale of 0.05
arcsec/pixel.

\subsection{Target selection for the medium-resolution
  spectroscopy}\label{sec_target_selection} The target selection for the
medium-resolution spectroscopy was based on the FDF spectroscopic survey (Noll
et al. \cite{noll2004}).  The FDF spectroscopic survey aimed at obtaining low-resolution spectra (R$\approx$200) of
intrinsically bright galaxies with a photometric redshift (Bender et al.
\cite{bender2001}; Gabasch et al. \cite{gabasch2004}) between $z \approx$ 1
and 5 with a high signal-to-noise ratio ($\geq$10).  Two different selection criteria were applied to select candidates for
the medium-resolution spectroscopy from the FDF spectroscopic catalog. First,
galaxies with strong \lya\ emission were selected. Second, galaxies with
bright UV-restframe continuum were added. We included only galaxies whose
spectral feature(s) of interest coincide with the wavelength range of the
FORS2 grisms 1400V and 1200R and whose expected signal-to-noise ratio was
sufficiently high.  For the 1400V (1200R) grism, UV-bright continuum galaxies
with $z = 2.3 - 3.5$ ($z = 3.0 - 3.5$) and \mg\ $ \leq\ $ 24.5 mag (\mR\ $
\leq\ $ 24.5 mag) were selected. Moreover, galaxies with a \lya\ emission
strength of \flya\ $\geq\ $ 30 (20) $\times$ \twwattm\ at $z = 3.0 - 3.5$ ($z
= 4.5 - 5$) were included.  We also added a few secondary targets selected by
their photometric redshifts. Some additional bright objects were included to
support the mask positioning during the observation.

\subsection{Observations \& data reduction of the medium-resolution sample}\label{sec:obs_datared}
The observations were obtained with FORS2 at the VLT UT4 using the holographic
grisms 1400V and 1200R. The spectral resolution was R $\approx$ 2000. The
spectral range of the 1400V (1200R) grism was about 4500 to 5800 (5700 to
7300) \AA . The central wavelengths depend on the target position in the focal
plane. All data were collected in service mode using one single MXU mask for
each grism.  The FORS2 detector, which consists of two 2k $\times$ 4k MIT
CCDs, was used in the 100kHz readout mode with high gain. A 2$\times$2 binning
was performed during the readout. The observations with the 1400V grism were
carried out during August 2002. Eight single exposures with each 47 min integration
time were taken, resulting in a total integration time of 6.25 h. The average
seeing was 0.81''. The observations with the 1200R grism were carried out
during July - September 2003. Fourteen single exposures with a total integration
time of 10.05 h were obtained. One exposure was excluded from the reduction
process, because of moonlight contamination. Therefore, the total effective
integration time was 9.45 h. The average seeing was 0.92''.

The raw data were reduced using the MIDAS-based FORS pipeline (Noll et al.
\cite{noll2004}; Tapken \cite{tapken2005}). The two-dimensional spectra were
flat-fielded with a dome flatfield and were wavelength-calibrated using
calibration spectra of gas discharge lamps. One-dimensional spectra were
extracted with a signal-to-noise optimising algorithm (Horne
\cite{horne1986}). A first flux calibration, which included the correction for
extinction by the atmosphere, was performed using spectra of standard stars
obtained during the same night.  The wavelength calibration was verified and,
if necessary, corrected by determining the position of sky lines. The
one-dimensional spectra were then co-added according to their weighted
signal-to-noise. The efficiency of holographic grisms strongly varies with the
angle of incidence and thus with the object's position in the telescope's
focal plane. Therefore, the low-resolution spectra were used to improve the
flux calibration.  Some objects were observed only on the medium-resolution spectra. In these cases the varying response function was corrected using the sky
spectrum. For this, the sky was extracted from all slits and was divided by
the sky in those slits, which had almost the same position as the
corresponding slits used in the standard star calibration exposures. In total,
43 objects have been reduced, and six objects were included in both setups. The
majority of the 43 objects are galaxies. Only one object is classified as a
quasar (QSO Q0103-260 or FDF-4683), and one as a star (FDF-0511).  For more
information on the 43 objects see Tapken (\cite{tapken2005}).

\section{Results}\label{sec_results}

Since the low-resolution spectra have a broader wavelength range and
consequently more spectral features, the redshift for all objects with
low-resolution spectra available was taken from the spectroscopic catalog of
Noll et al. (\cite{noll2004}). FDF-1267 coincided by chance with the slit of a
primary target and, therefore, has no corresponding low-resolution spectra.
The medium-resolution spectrum shows a strong emission line that has been
identified as likely \lya\ emission, due to the non-detection of other
emission lines and the asymmetric line profile. The redshift of this object
is derived from the \lya\ emission line. FDF-8304 was selected because its
spectral features were expected to coincide with the spectral range of the
grism based on the photometric redshift of \zphot\ = 4.02.  The
medium-resolution spectrum shows a \lya\ emission line and several low
ionisation interstellar absorption lines. In this case again the redshift is
derived from the \lya\ line.

Thirty galaxies of the medium-resolution sample have a redshift of $z >$
2. Twelve of these 30 galaxies having a sufficiently high continuum SNR for studying their
absorption line spectra as discussed in Mehlert et al. (\cite{mehlert2006}).
In the present paper the analysis of the medium-resolution spectra is
restricted to  16 of the 30 galaxies, where (a) the medium-resolution
spectrum contains the \lya\ line, either in emission or absorption and where
(b) the galaxy is included in the FDF photometric catalog (Heidt et al.
\cite{heidt2003}). Their redshift, apparent magnitude \mR , spectral coverage,
and average continuum SNR are listed in Table \ref{def_lya1}. These 16
galaxies are referred to, in the following, as the \lya\ medium-resolution
sample. This sample covers a redshift range of $z \approx$ 2.7 to 5.
\begin{table*}
\caption{The  \lya\ medium-resolution sample.  The ID corresponds to the
  catalogs of Heidt et al. (\cite{heidt2003}) and Noll et
  al. (\cite{noll2004}). The redshift $z$ is from Noll et
  al. (\cite{noll2004}), 
  except for FDF-1267 and FDF-8304, where the redshift is derived 
from medium-resolution spectra. The 2`` aperture magnitude  \mR\  of the
  galaxies is taken from Heidt et al. (\cite{heidt2003}). The SNR is the average continuum SNR measured in the complete
  wavelength range of the spectra. The covered spectral range is given for the
  restframe of the corresponding galaxies. }
   
\label{def_lya1}
\vspace{2mm}
\centering
\begin{tabular}{|c|c|c|c|c|c|c|}
\hline
ID & z &  \mR & Spectral range&  Spectral range &SNR & SNR  \\
   &   & [mag]&  1400V &            1200R       &1400V & 1200R   \\
\hline
1267& 2.788 $\pm$ 0.001 & 27.08 $\pm$ 0.07 & 1158 - 1465 & - &  0.36 &  -  \\
1337& 3.403 $\pm$ 0.004 & 24.15 $\pm$ 0.01 &996 - 1261 & 1235 - 1569&  3.53 & 4.72 \\
2384& 3.314 $\pm$ 0.004 & 24.60 $\pm$ 0.01 & 1018 - 1317& 1301 -
1643 & 1.17 & 3.01  \\
3389& 4.583 $\pm$ 0.006 & 25.56 $\pm$ 0.02 & - &1097 - 1368 &   -   & 0.71  \\
4454& 3.085 $\pm$ 0.004 & 26.10 $\pm$ 0.03 &  1070 - 1387& - & 0.65 &  -        \\
4691& 3.304 $\pm$ 0.004 & 24.79 $\pm$ 0.01 &  1089 - 1393&- &  3.78 &  -     \\
5215& 3.148 $\pm$ 0.004 & 24.53 $\pm$ 0.01  & 1062 - 1373& 1468 - 1832 &  3.01 & 3.98  \\
5550& 3.383 $\pm$ 0.004 & 23.95 $\pm$ 0.01  &1053 - 1351& 1315 - 1656&  3.78 & 6.47 \\
5744& 3.401 $\pm$ 0.003 & 24.81 $\pm$ 0.01  & 1022 - 1317& - & 2.05 &  -  \\
5812& 4.995 $\pm$ 0.006 & 27.55 $\pm$ 0.11  & - & 1040 - 1294&   - &  0.69 \\
5903& 2.774 $\pm$ 0.003 & 23.13 $\pm$ 0.01  & 1181 - 1523& - &  12.27& -   \\
6063& 3.397 $\pm$ 0.004 & 23.37 $\pm$ 0.01  & 1087 - 1384& 1273 -
1609 &  4.18&5.83\\
6557& 4.682 $\pm$ 0.006 & 25.94 $\pm$ 0.02  & -  & 1003 - 1265 &    -  &0.69\\
7539& 3.287 $\pm$ 0.003 & 24.10 $\pm$ 0.01  &1122 - 1428& 1351 - 1698 &  3.45&4.88 \\
7683& 3.781 $\pm$ 0.004 & 24.92 $\pm$ 0.01  & - & 1199 - 1510 & -   &2.2 \\
8304& 4.205 $\pm$ 0.003 & 24.98 $\pm$ 0.01  & - & 1145 - 1433 &  -   &3.37  \\
\hline
\end{tabular}
\end{table*}
  
In addition to \lya\ the galaxies show spectral features such as interstellar
absorption lines, stellar wind lines, photospheric lines, and several nebular
emission lines.  For a detailed discussion of the absorption
features, we refer to Mehlert et al. (\cite{mehlert2006}) and Noll et al.
(\cite{noll2004}).  All objects of the \lya\ medium-resolution sample, which
were included in the FDF spectroscopic survey, are classified as starburst
galaxies by Noll et al. (\cite{noll2004}), based on the observed UV
properties. The spectral energy distribution, the stellar wind lines, and the
interstellar absorption line of these objects are typical of starburst
galaxies. Furthermore, the line ratio of the nebular lines are typical of
starburst galaxies. We could not detect broad lines in either the
low-resolution spectra or the medium-resolution spectra, which would indicate
AGN activity.

\subsection{Observed Ly$\alpha$ line properties}

\subsubsection{Fluxes and equivalent widths}\label{sec_lya_fluxew} 
The \lya\ profiles often display both an absorption and an emission
component. Only the spectral resolution of the medium-resolution spectra is
 high enough to separate the two components for each spectrum.  We
measured the emission line fluxes in the medium-resolution spectra
interactively with MIDAS. To calculate the equivalent widths, we estimated the
continuum at the wavelength of \lya\ from the continuum redwards of the \lya\ 
line.  The observed \lya\ emission fluxes \flya\ and equivalent widths of the
emission component \ewlyaem\ derived for the \lya\ medium-resolution sample
are given in Table \ref{def_lya2}.  The error given in Table \ref{def_lya2}
includes only the contribution of the noise. Any additional error caused by a
possibly wrong continuum definition is not included. We also measured the
total equivalent widths \ewlya , which includes the emission and the
absorption component.  However, the corresponding measurements are less
accurate because of the underlying IGM absorption contributions. In this paper
the equivalent width is defined as positive for emission lines. All equivalent
widths are given in the restframe of the corresponding galaxy.  Except for
FDF-6063, all objects of the medium-resolution sample show a \lya\ emission
component. FDF-6063 shows only a broad absorption (\ewlya\ = -17.1 $\pm$ 0.9
\AA ).  The measured \lya\ fluxes reach up to 186 $\times$ \twwattm . For
strong \lya\ emitters, the emission component dominates the total \lya\ line.
Some galaxies show a strong absorption component and a weak \lya\ emission
line. The total \lya\ emission line is dominated in these cases by the
absorption component.  An example is FDF-5903 (\ewlya\ = -12.0 $\pm$ 0.5 \AA ,
\ewlyaem\ = 0.6 $\pm$ 0.1 \AA ).  Eight of the 16 galaxies of the
medium-resolution sample have total equivalent width \ewlya\ higher than 20
\AA . We also analysed the \lya\ emission line of 91 low-resolution spectra of
high-redshift galaxies in the FDF spectroscopic catalog.  The total \lya\ 
equivalent widths range from absorption (\ewlya $\approx$ -20 \AA ) to strong
emission (\ewlya\ $\approx$ 200 \AA ), while most galaxies have \ewlya\ of $<$
20 \AA .

\begin{table*}
\caption{Properties of the  \lya\ medium-resolution sample.  The continuum
 slope $\beta$  is taken from Noll et al. (\cite{noll2004}).  They
  measured the continuum slope $\beta$ in the low-resolution spectra of the
  FDF spectroscopic survey following  Leitherer et al. (\cite{leitherer2002}). The \lya\ fluxes \flya\
  and equivalent widths \ewlya\  refer only to the emission component
  (Sect. \ref{sec_lya_fluxew}). No \lya\ emission line was detected in the
  medium-resolution spectra of FDF-6063.  The star-formation rates \sfruv\
   and  \sfrlya\ were derived from the UV flux, respectively \lya\ flux, using
  the calibration of Kennicutt (\cite{kennicutt1998}) and assuming
  Case B recombination.   
  The  line width $FWHM$(\lya ) refers to the complete profile (Sect \ref{sec_obs_profile}) . }
\label{def_lya2}
\vspace{2mm}
\centering
\begin{tabular}{|c|c|c|c|c|c|c|c|c|}
\hline
ID & z & \sfruv [\Msyr ] & $\beta$ &   \flya
[\twwattm ] &\sfrlya [\Msyr ] & \ewlyaem\ [\AA ] & FWHM(\lya ) [km/s]  \\
\hline
1267& 2.788 $\pm$ 0.001 & 1.16 $\pm$ 0.25&- &25.38 $\pm$ 1.30 &  1.49 $\pm$ 0.08& 
 129.8 $\pm$ 27.41 & 235 $\pm$ 32 \\
1337& 3.403 $\pm$ 0.004 & 27.28 $\pm$ 1.15 & -2.43 &22.09 $\pm$ 1.52 &2.10 $\pm$ 0.14  &6.69 $\pm$ 0.46 & 597 $\pm$ 84\\
2384& 3.314 $\pm$ 0.004 & 22.74 $\pm$ 0.77 & -0.55 &121.08 $\pm$
3.01 & 10.8 $\pm$ 0.27 &83.19 $\pm$  3.89 & 283 $\pm$ 47\\
3389& 4.583 $\pm$ 0.006 &14.85 $\pm$ 2.47  &- & 46.51 $\pm$ 2.23
& 9.2 $\pm$ 0.38 & 38.82 $\pm$ 10.95 & 354 $\pm$ 70 \\
4454& 3.085 $\pm$ 0.004 & 1.98 $\pm$ 0.49 &-2.42 &29.91 $\pm$
1.02 &2.25 $\pm$ 0.08 & 74.38 $\pm$ 11.84 & 323 $\pm$ 47 \\
4691& 3.304 $\pm$ 0.004 &17.88 $\pm$  0.75  &-2.46 &184.33 $\pm$
1.61 &16.31 $\pm$ 0.14 & 79.44 $\pm$ 1.61 & 840 $\pm$ 115 \\
5215& 3.148 $\pm$ 0.004 & 26.20  $\pm$ 0.80  &-1.71 & 121.55
$\pm$ 2.63&  9.57 $\pm$ 0.21&  32.48 $\pm$ 1.06 & 483 $\pm$ 90 \\
5550& 3.383 $\pm$ 0.004 & 44.78  $\pm$ 1.07 & -1.81& 34.89 $\pm$
2.14 &  3.27 $\pm$ 0.2 & 6.36 $\pm$ 0.40 & 424 $\pm$ 85\\
5744& 3.401 $\pm$ 0.003 & 21.23  $\pm$   0.87 & -1.02&  4.63$\pm$
1.06 &  0.44 $\pm$ 0.10& 3.30$\pm$ 0.77 & - \\
5812& 4.995 $\pm$ 0.006 &  5.24  $\pm$   0.79  &- &40.83 $\pm$ 0.78
&9.60 $\pm$ 0.18  & 153.8$\pm$ 26.6 & 226 $\pm$ 23 \\
5903& 2.774 $\pm$ 0.003 &63.14 $\pm$ 0.75  &-1.15&6.73 $\pm$
1.53 &  0.39 $\pm$ 0.09& 0.60  $\pm$ 0.14 & 627 $\pm$ 140\\
6063& 3.397 $\pm$ 0.004 &56.61 $\pm$ 1.28 &-2.02 & -&- &- &- \\
6557& 4.682 $\pm$ 0.006 &13.85 $\pm$ 1.39  &- & 16.57 $\pm$ 0.73 
&  3.35 $\pm$ 0.15 & 30.51 $\pm$ 3.04 & 380 $\pm$ 135 \\
7539& 3.287 $\pm$ 0.003 & 29.87 $\pm$ 0.78 &-1.74 & 28.01 $\pm$
1.84& 2.45 $\pm$ 0.46 & 6.84 $\pm$ 0.46   & 1430 $\pm$ 230\\
7683& 3.781 $\pm$ 0.004 &20.40 $\pm$ 1.46  & -1.16& 15.18 $\pm$
1.89& 1.85 $\pm$ 0.23 &  9.08 $\pm$ 1.16   & 435 $\pm$ 70 \\
8304& 4.205 $\pm$ 0.003 & 24.94 $\pm$ 5.48  &- &5.63 $\pm$ 0.74 &
0.88 $\pm$ 0.12  & 3.21 $\pm$ 0.43  & 500 $\pm$ 70 \\
\hline
\end{tabular}
\end{table*}

\subsubsection{Ly$\alpha $ line profiles}\label{sec_obs_profile}
\begin{figure*}
  \centering
  \subfigure[FDF-1267]{\includegraphics[width=2.5cm,angle=-90]{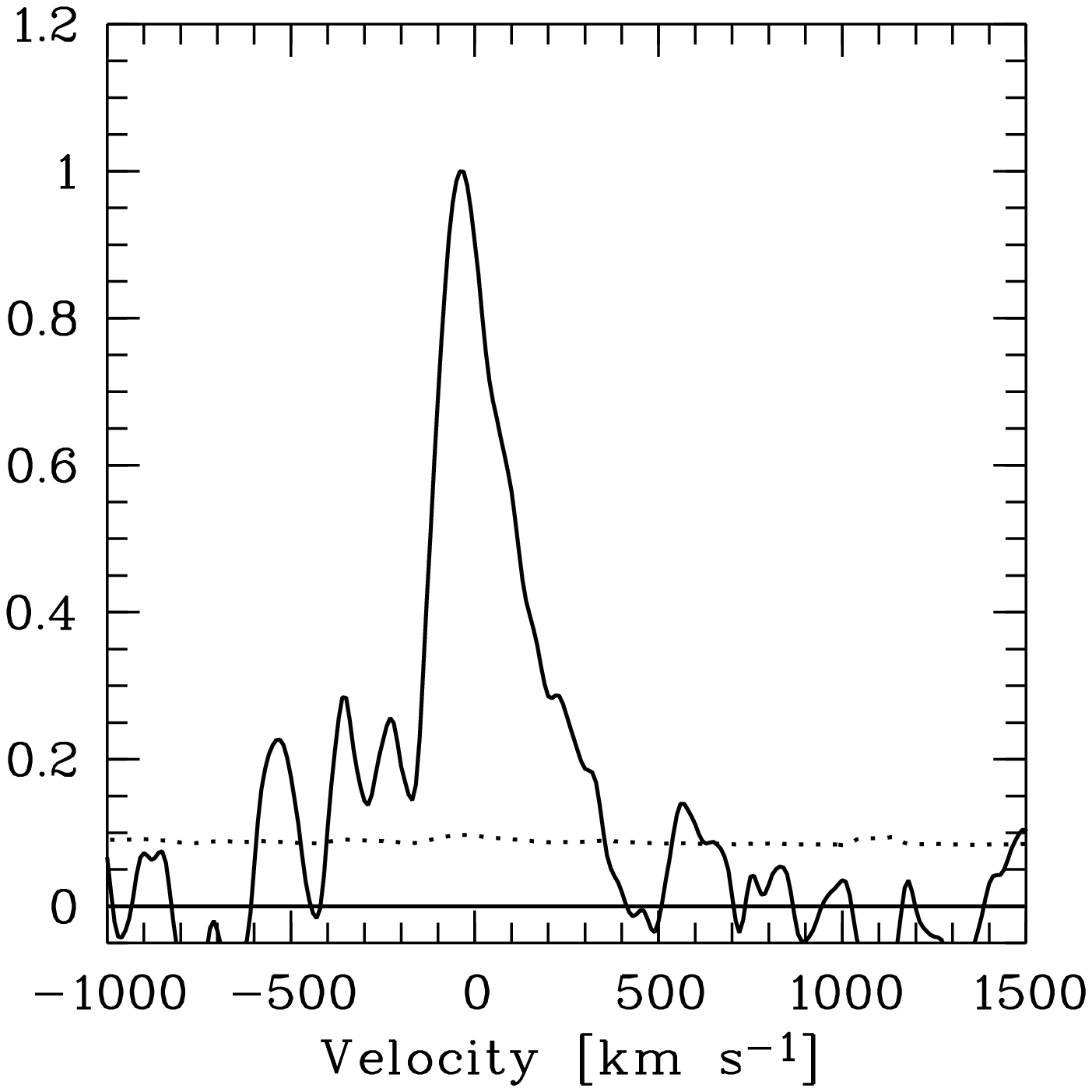}}
  \subfigure[FDF-1337]{\includegraphics[width=2.5cm,angle=-90]{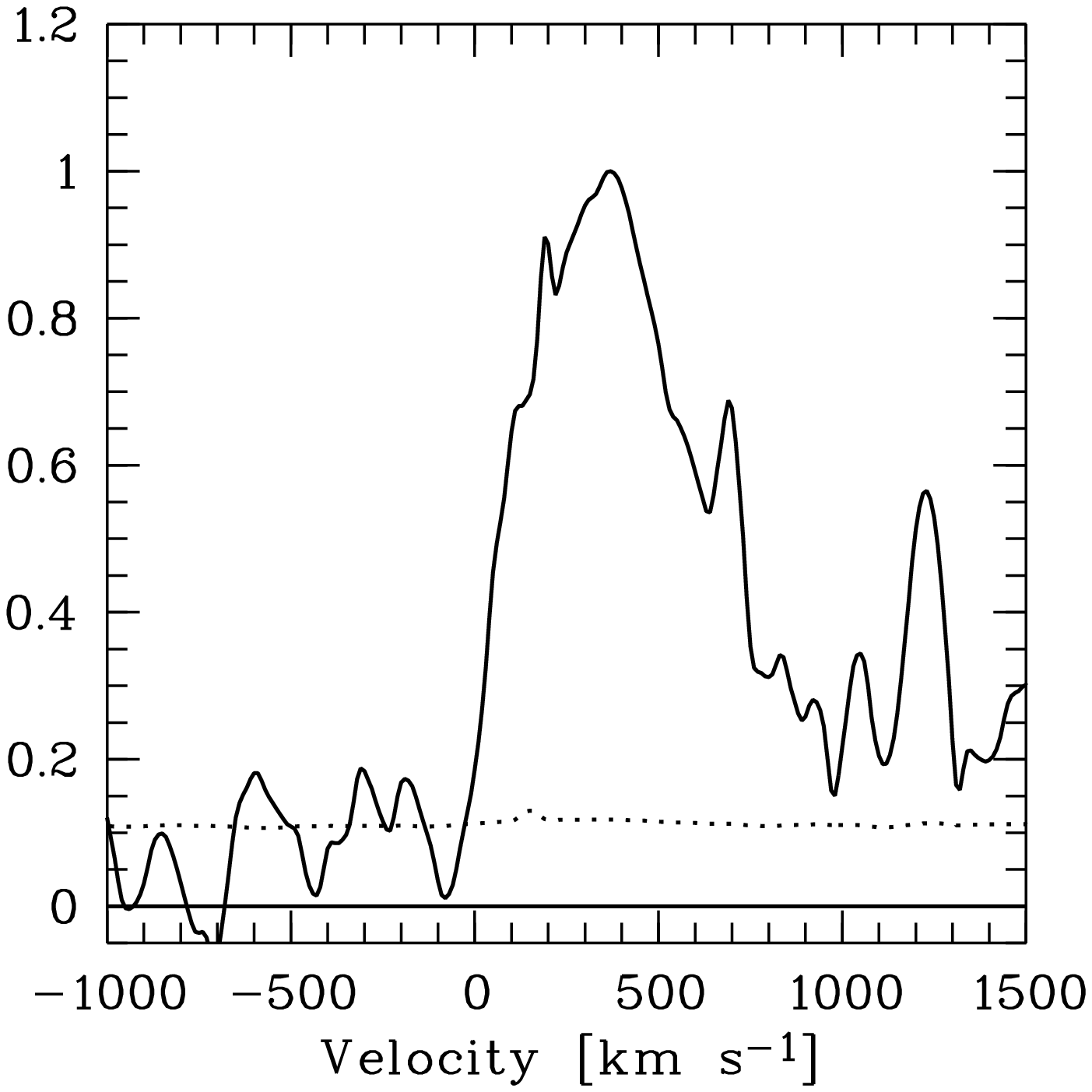}}
  \subfigure[FDF-2384]{\includegraphics[width=2.5cm,angle=-90]{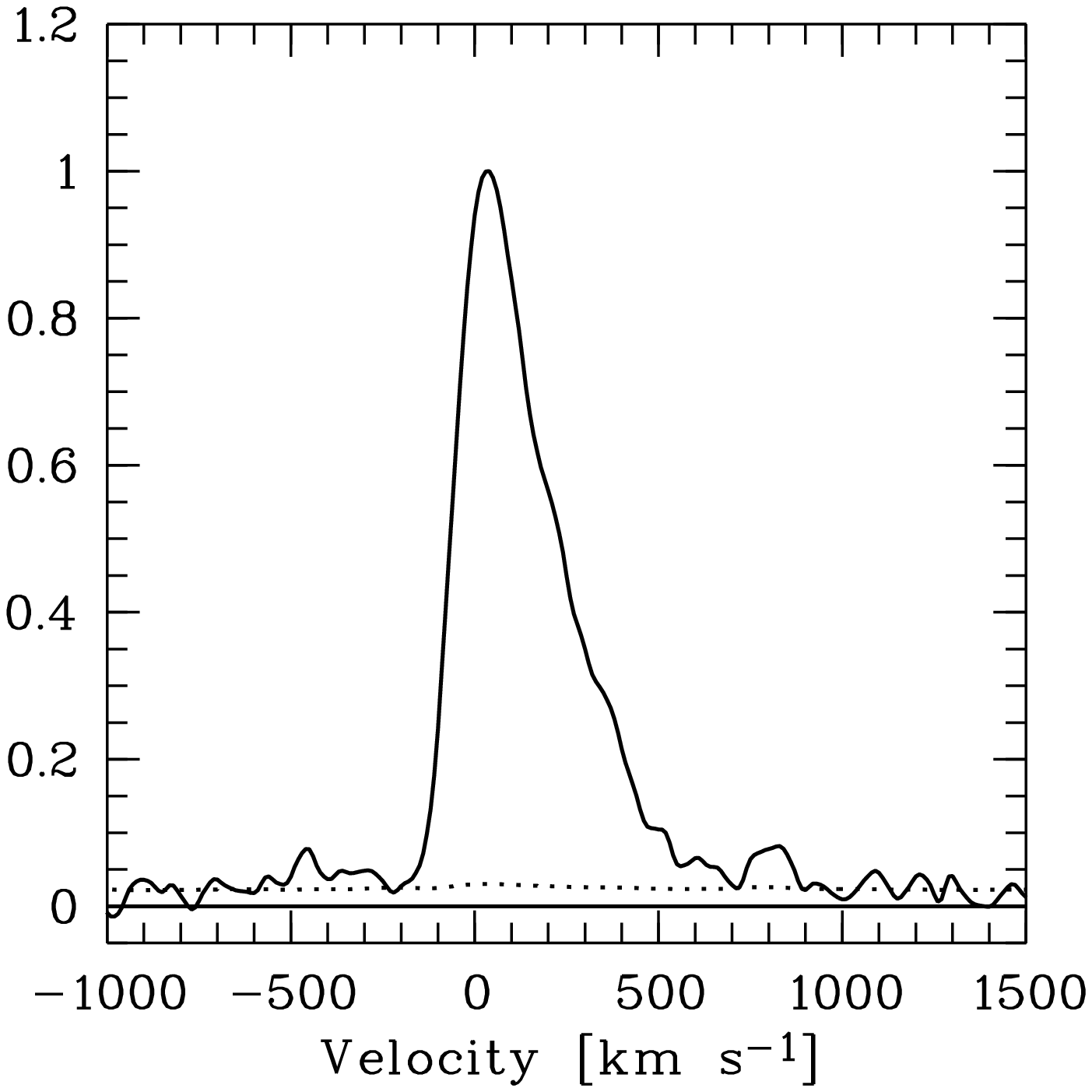}}
  \subfigure[FDF-3389]{\includegraphics[width=2.5cm,angle=-90]{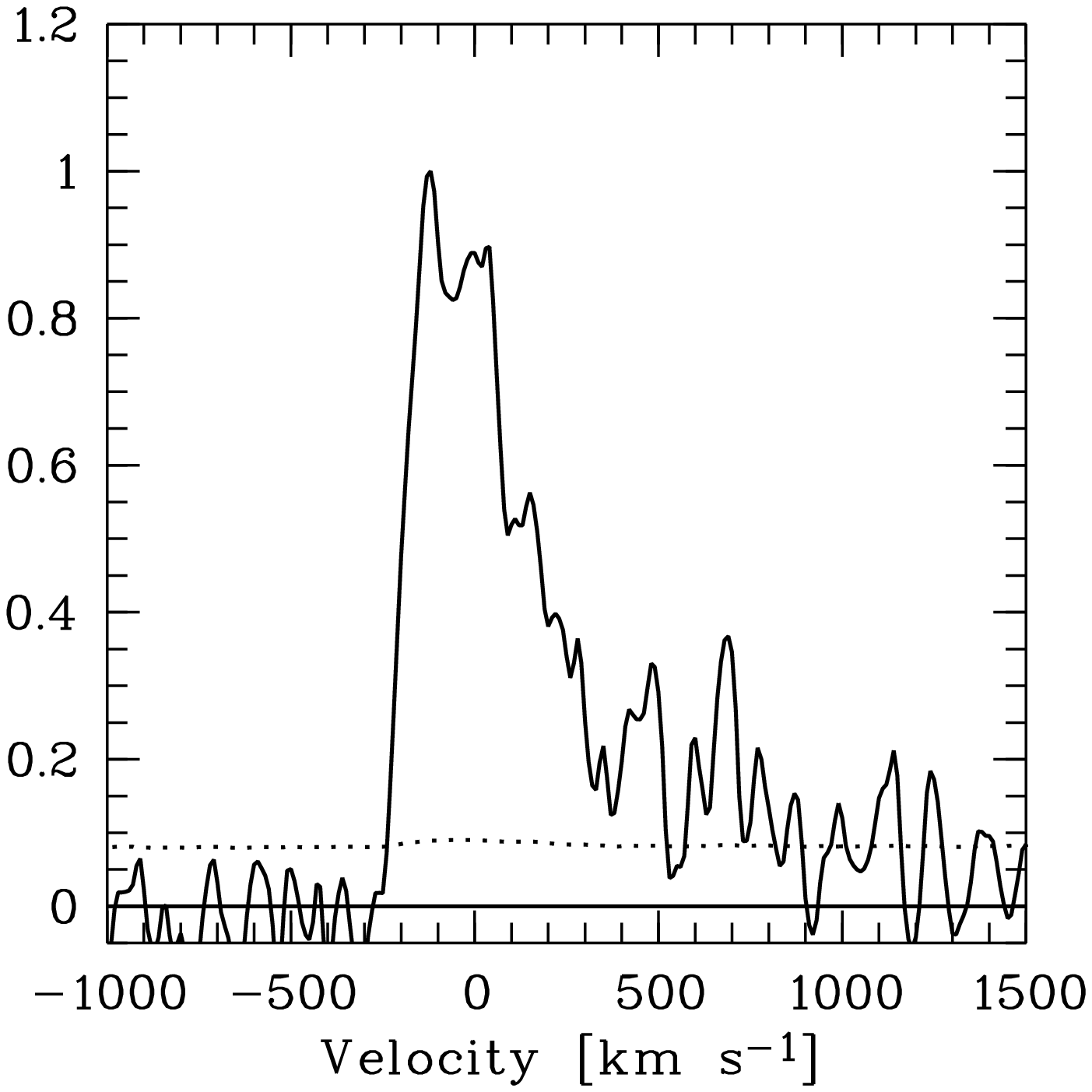}}
  \subfigure[FDF-4454]{\includegraphics[width=2.5cm,angle=-90]{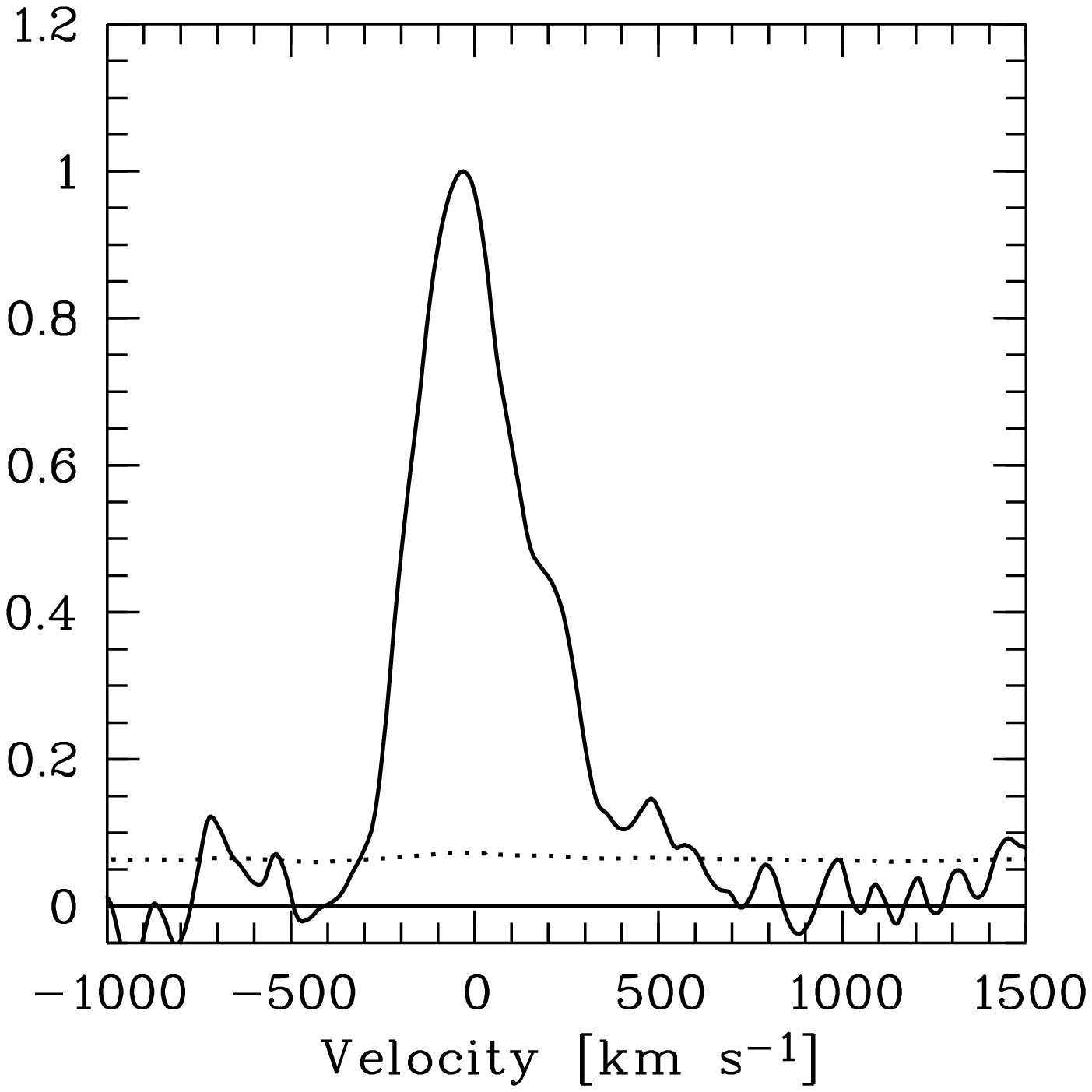}}
  \subfigure[FDF-4691]{\includegraphics[width=2.5cm,angle=-90]{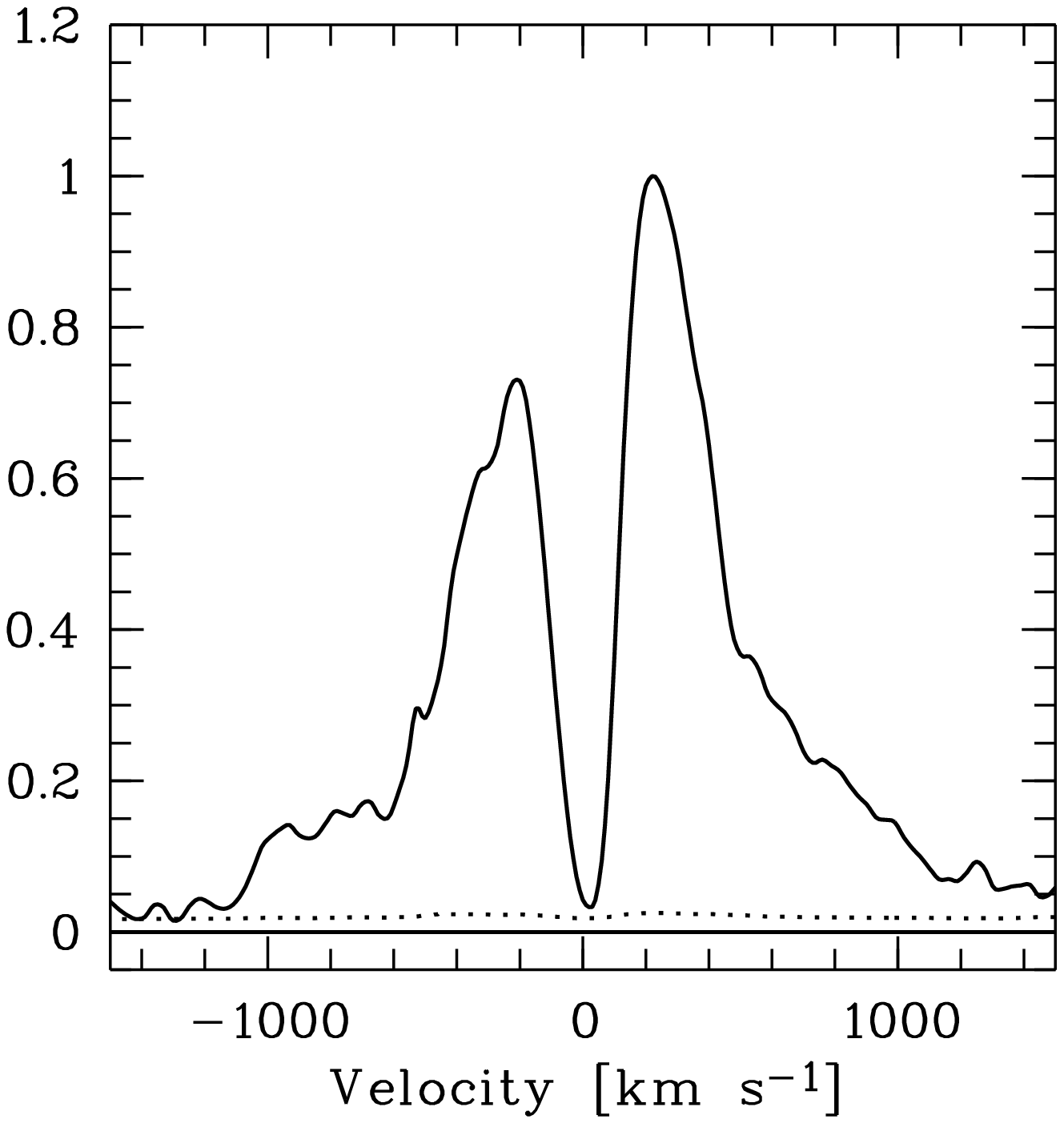}}
  \subfigure[FDF-5215]{\includegraphics[width=2.5cm,angle=-90]{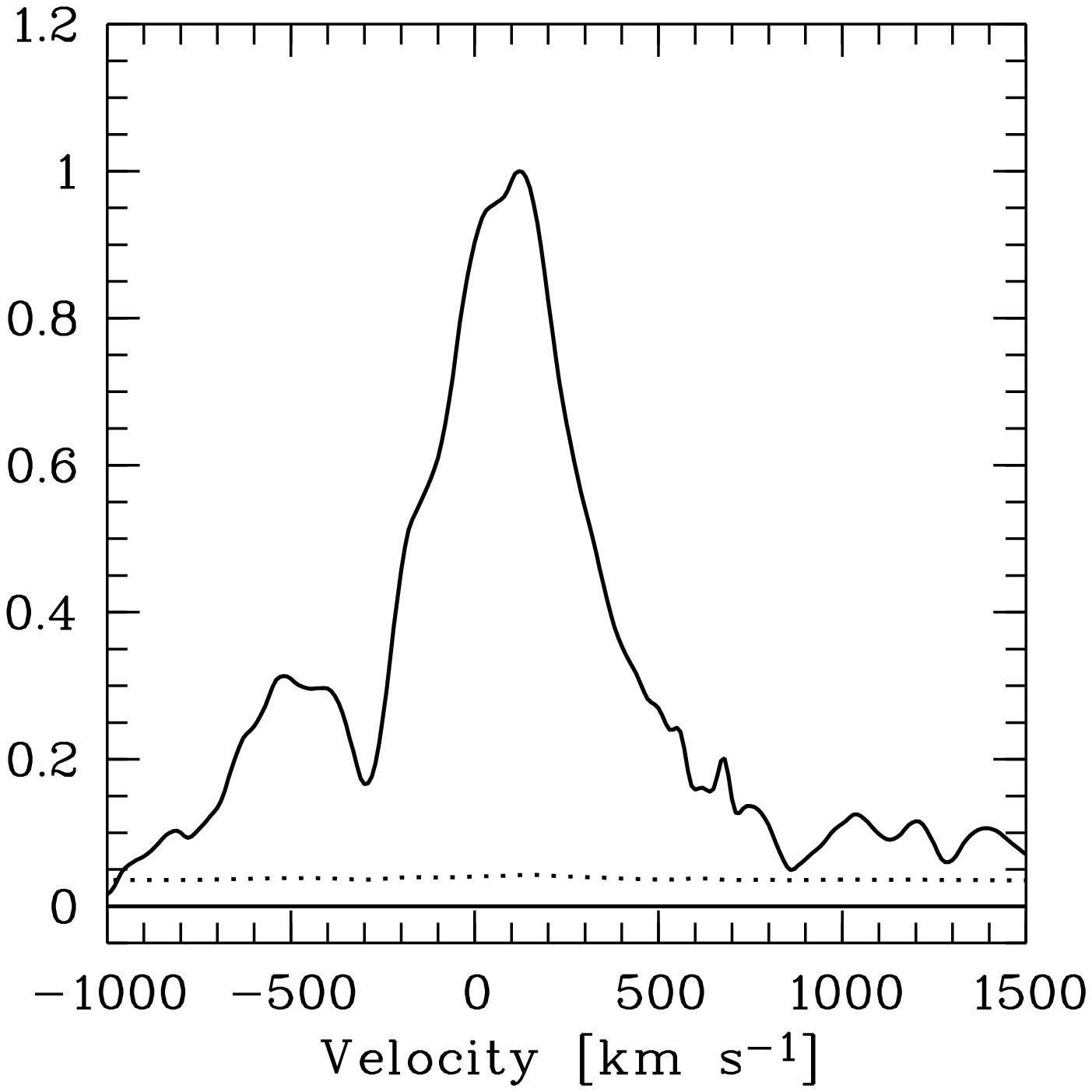}}
  \subfigure[FDF-5550]{\includegraphics[width=2.5cm,angle=-90]{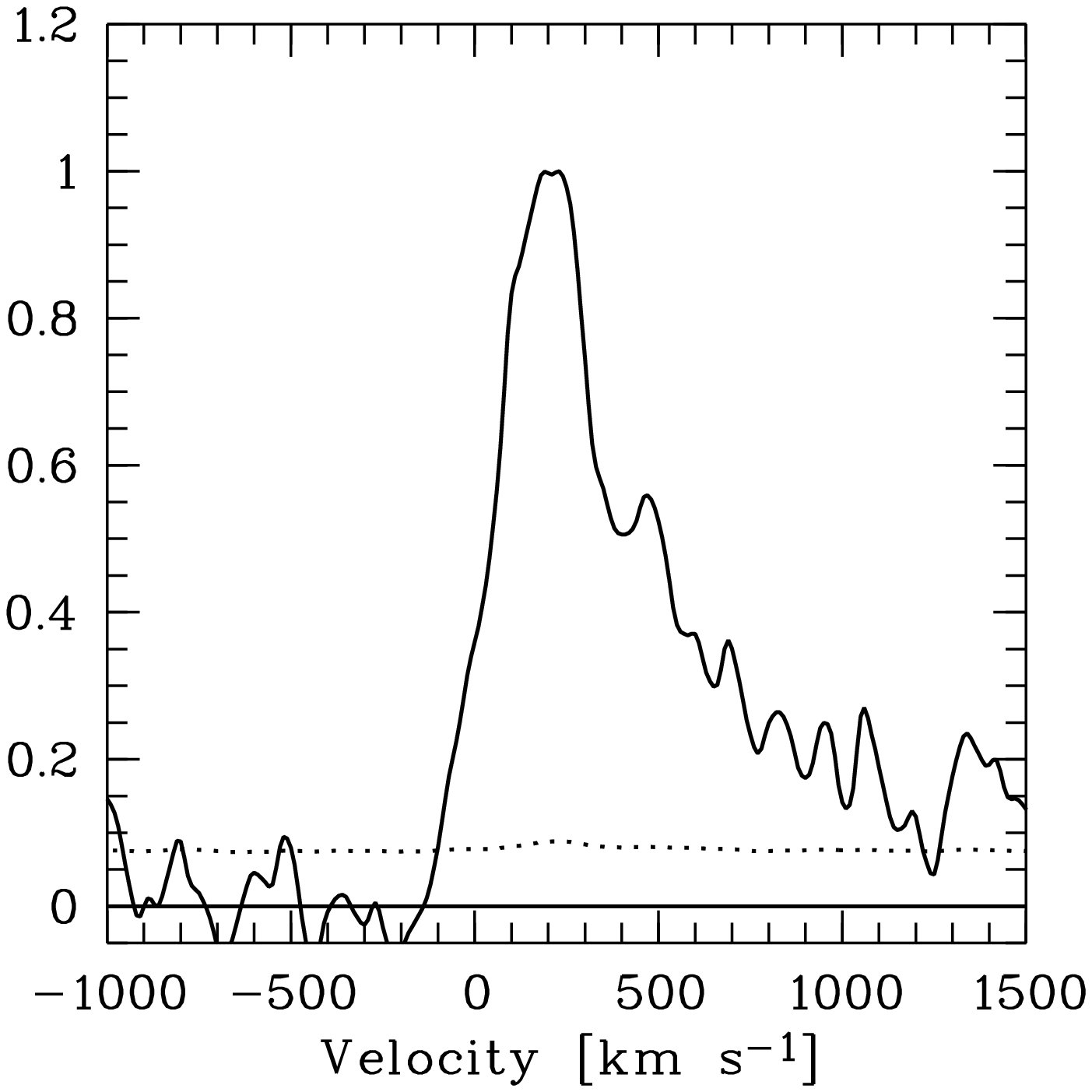}}
  \subfigure[FDF-5744]{\includegraphics[width=2.5cm,angle=-90]{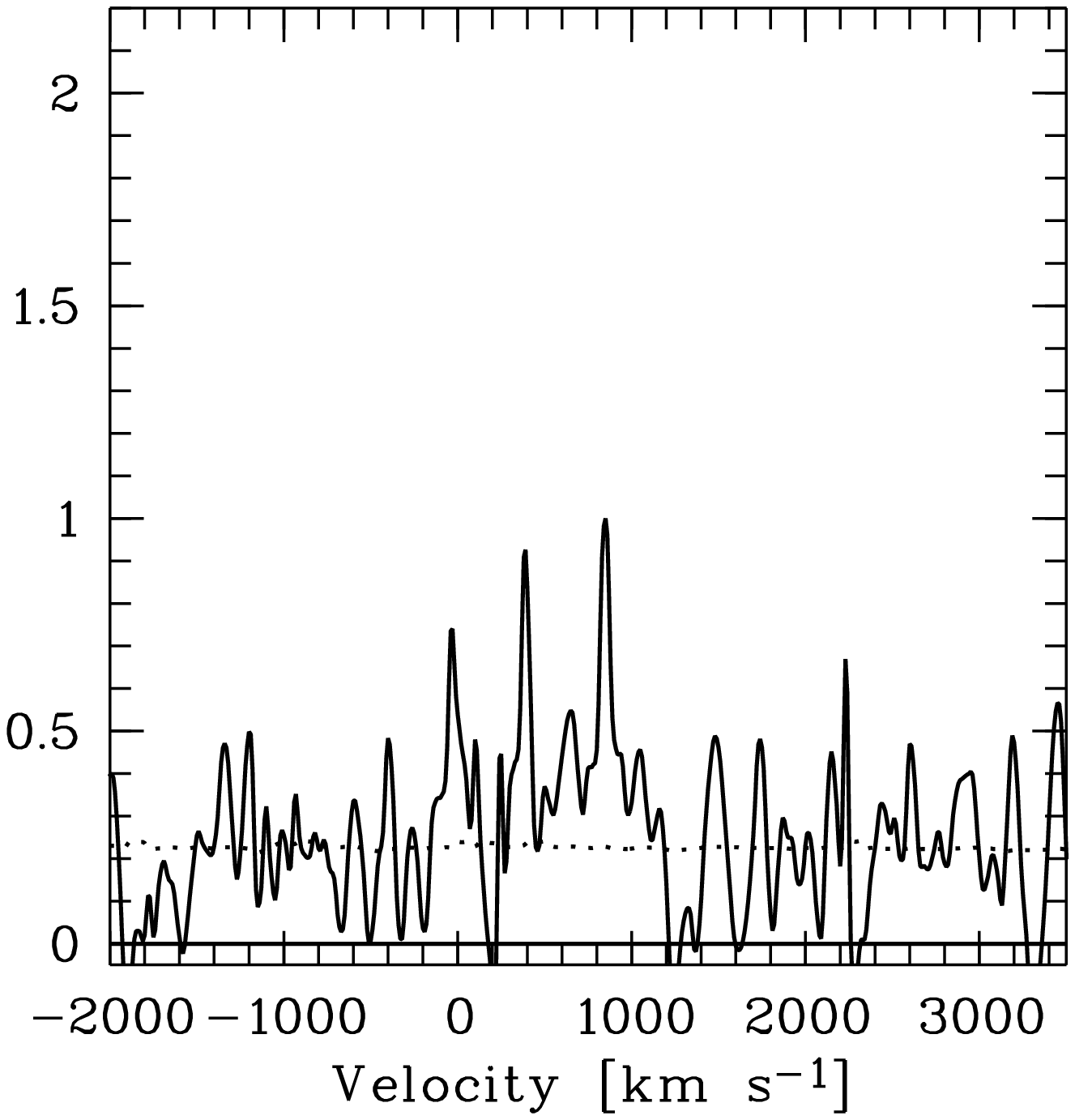}}
  \subfigure[FDF-5812]{\includegraphics[width=2.5cm,angle=-90]{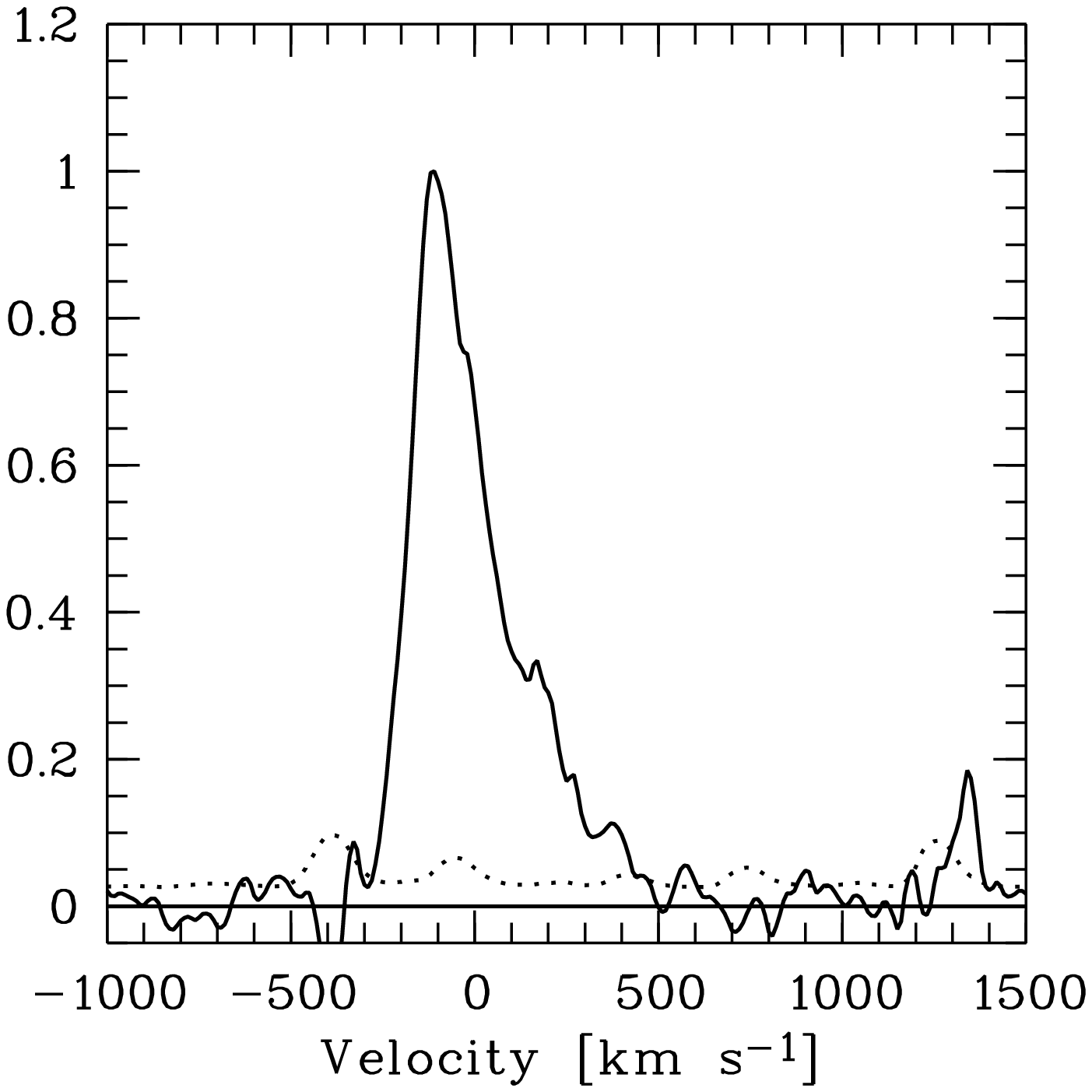}}
  \subfigure[FDF-5903]{\includegraphics[width=2.5cm,angle=-90]{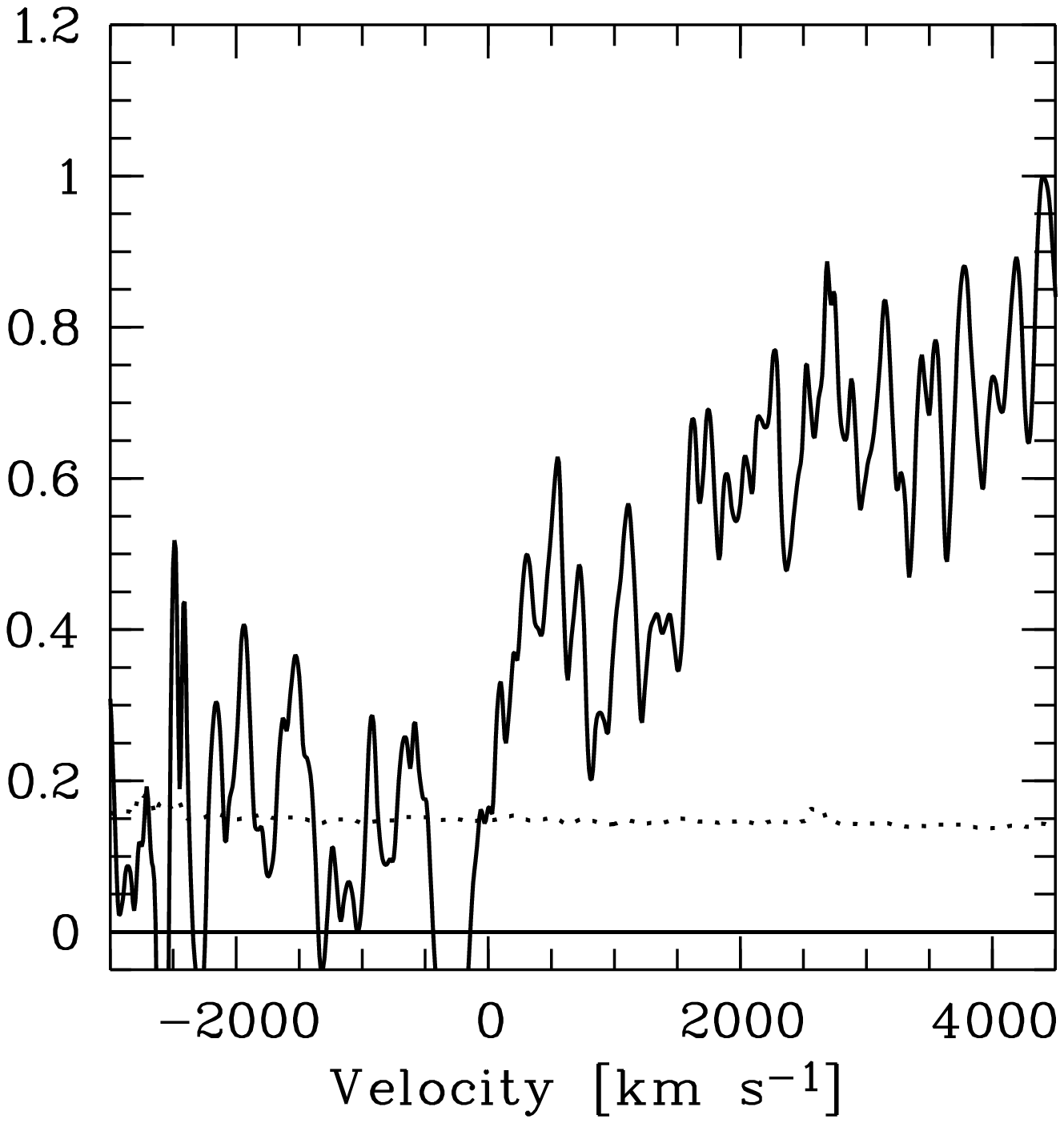}}
  \subfigure[FDF-6063]{\includegraphics[width=2.5cm,angle=-90]{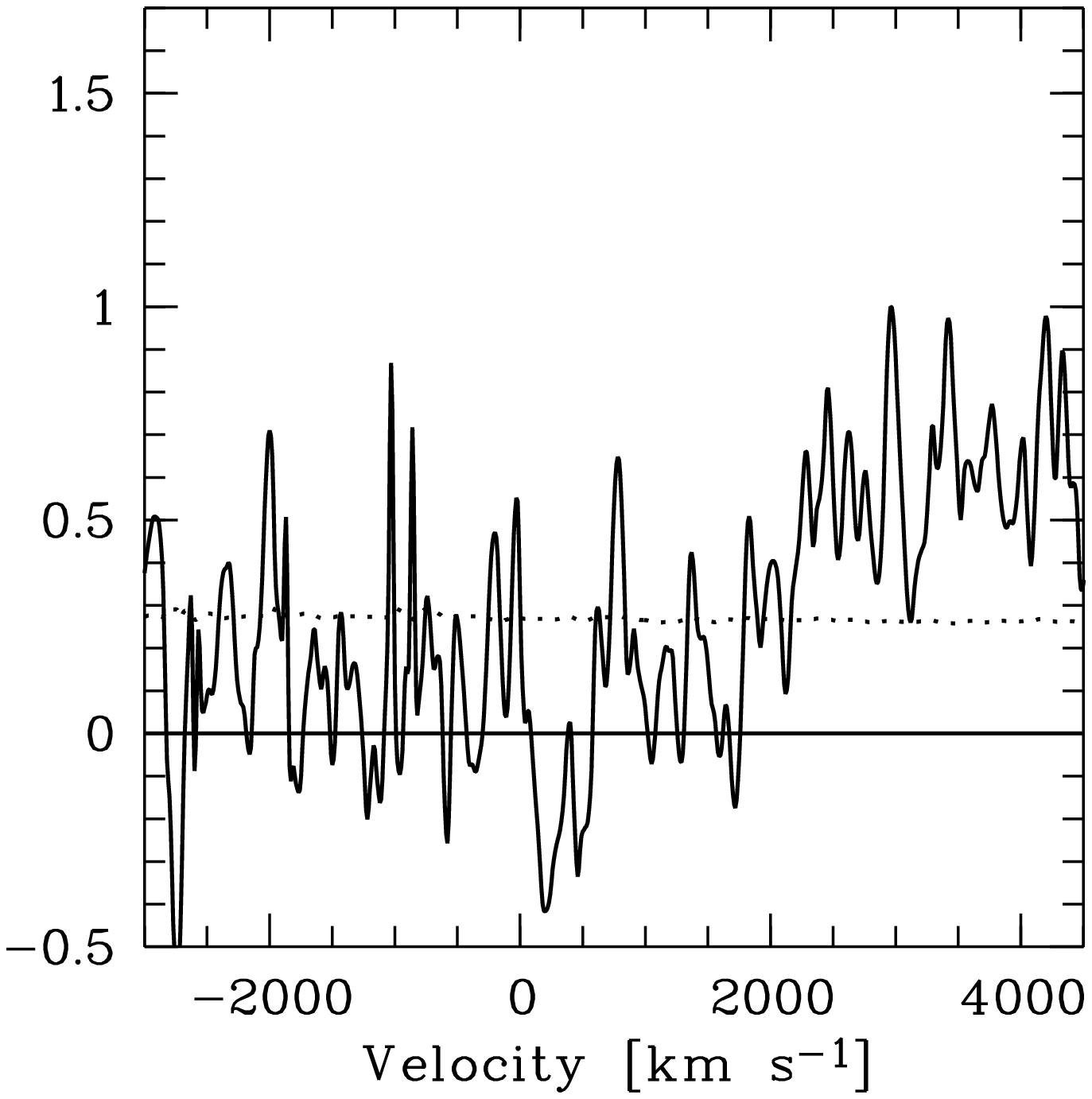}}
  \subfigure[FDF-6557]{\includegraphics[width=2.5cm,angle=-90]{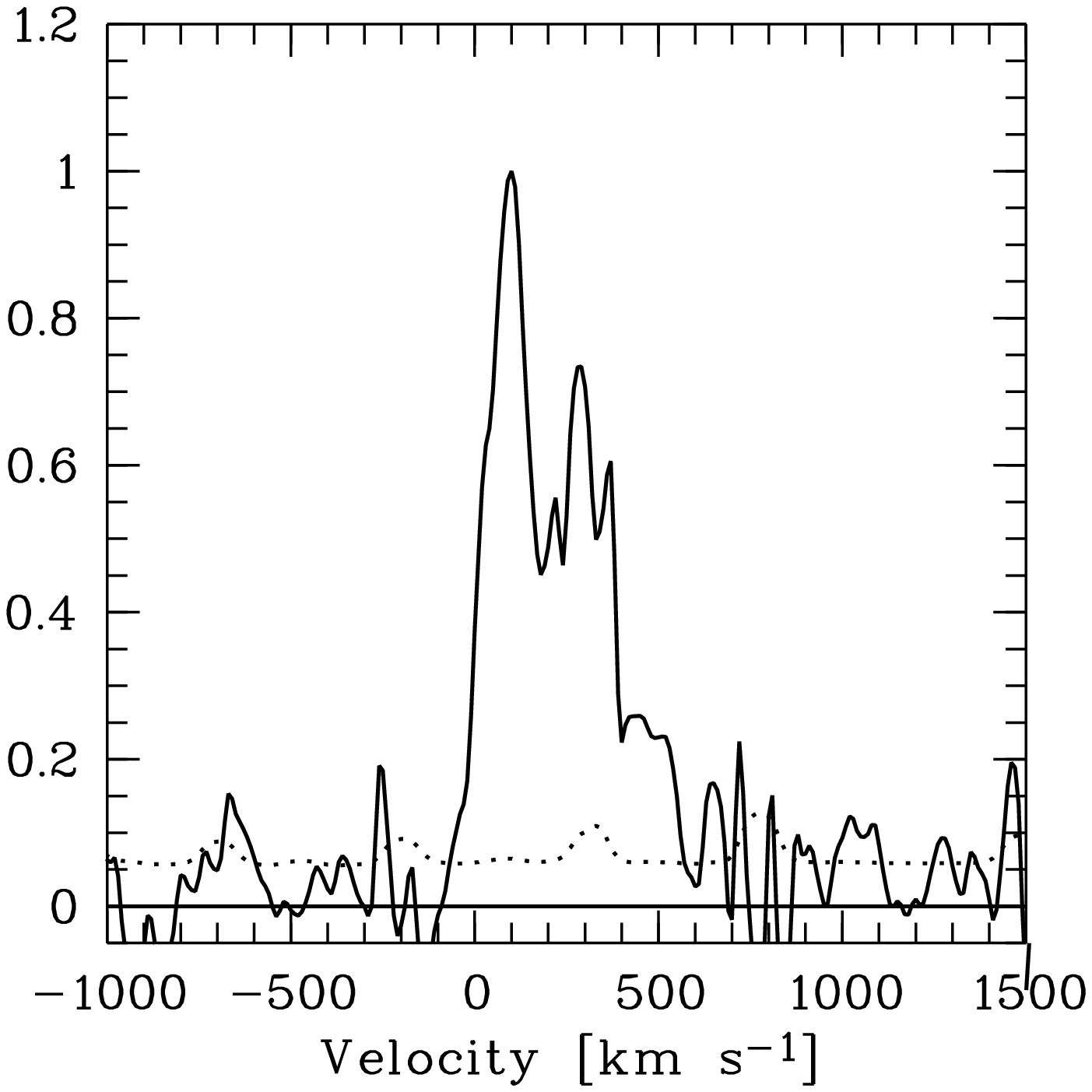}}
  \subfigure[FDF-7539]{\includegraphics[width=2.5cm,angle=-90]{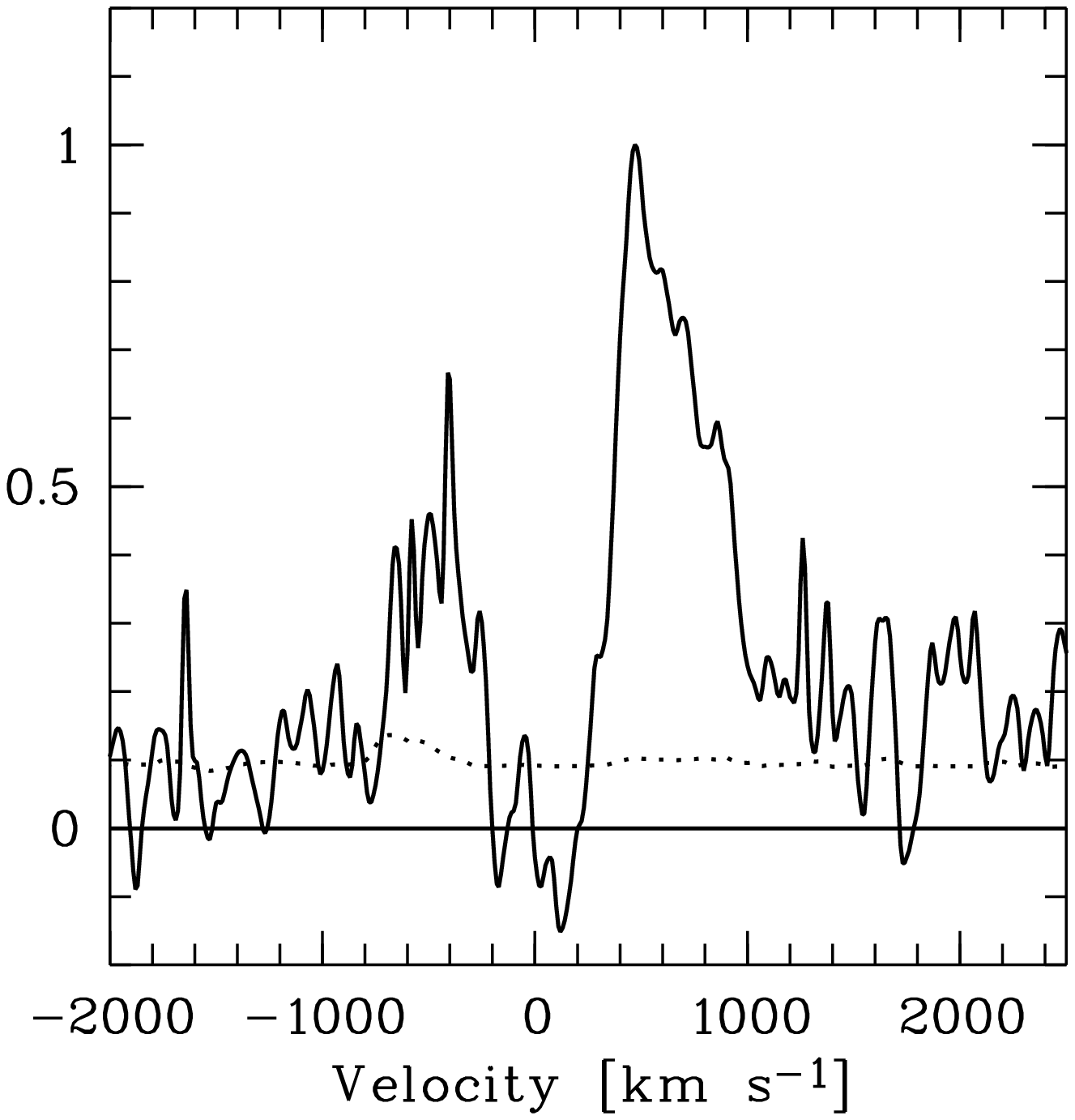}}
  \subfigure[FDF-7683]{\includegraphics[width=2.5cm,angle=-90]{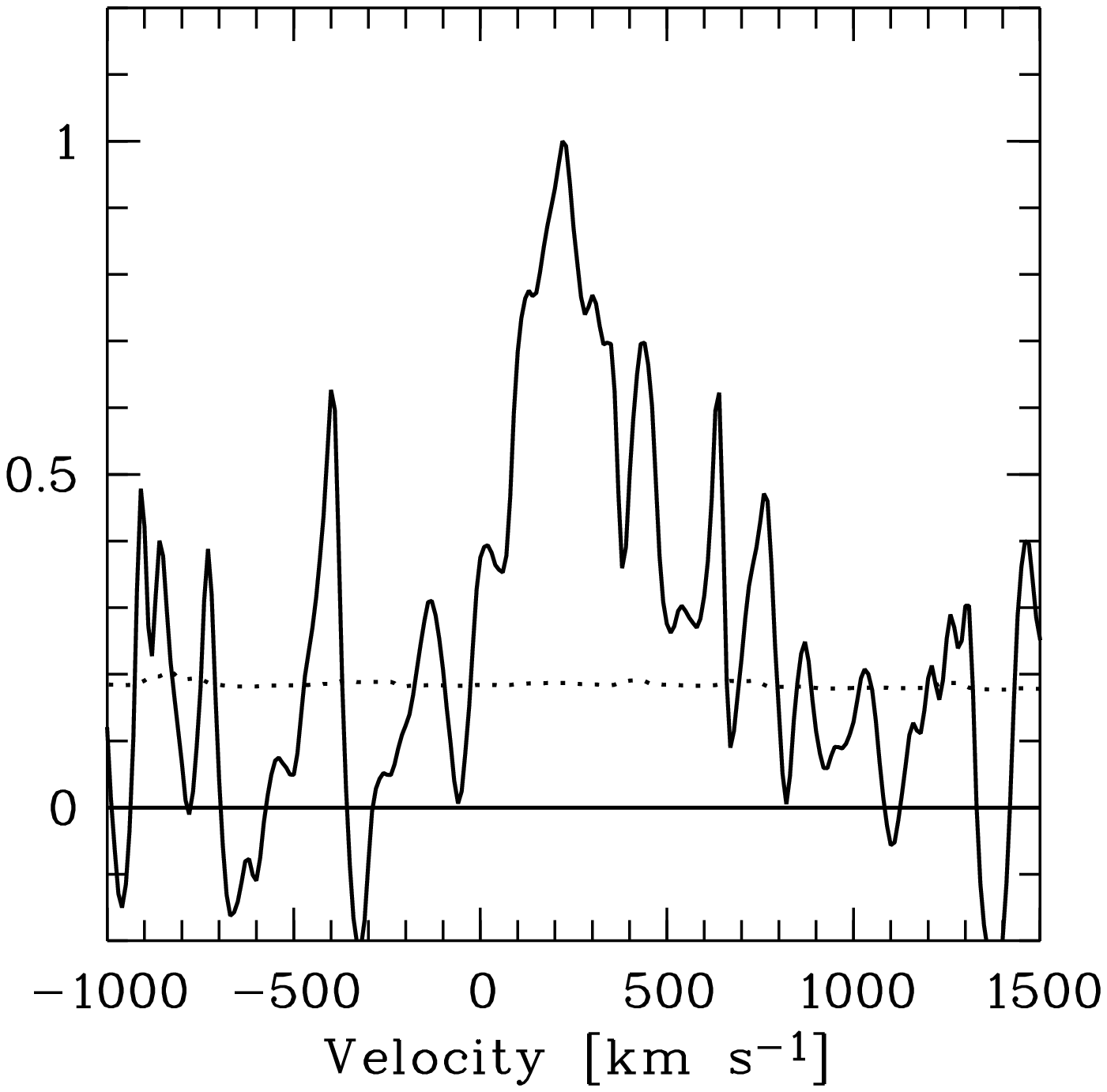}}
  \subfigure[FDF-8304]{\includegraphics[width=2.5cm,angle=-90]{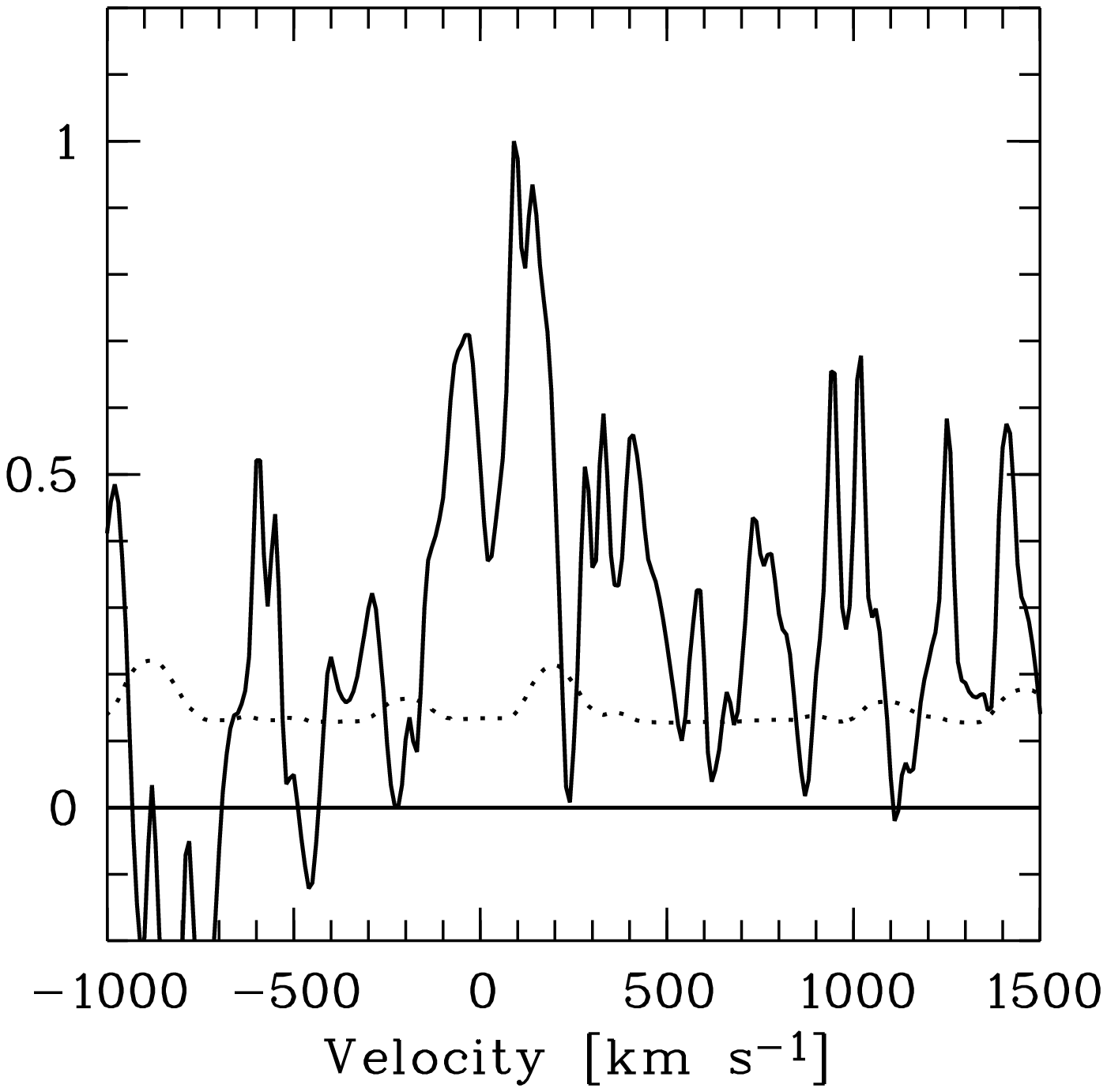}}
\caption{The observed medium-resolution spectra of the \lya\
  line. The abscissa for each figure gives the radial velocity relative to the
  redshift of the galaxies (see Table \ref{def_lya2}). The ordinate gives a
  normalised flux. The dotted line indicates the noise level.  }
\label{fig_profiles}  
\end{figure*}

In Fig. \ref{fig_profiles} all \lya\ profiles of the medium-resolution sample
are presented. The \lya\ profile displays a wide variety of morphologies, from
absorption (e.g., FDF-6063), to absorption with a weak emission (FDF-5903),
and then to
strong emission (e.g., FDF-2384). Most profiles show one asymmetric emission
line with a sharp drop on the blue side and a pronounced red wing (e.g.,
FDF-2384, FDF-5550). This profile type is referred to hereafter as the
``asymmetric line profile''.  FDF-4691, FDF-5215, and FDF-7539 show a
secondary emission line blueshifted with respect to the main emission line.
This profile type is called ``double-peak profile'' hereafter.
 
\begin{figure}
  \includegraphics[angle=-90,width=9cm]{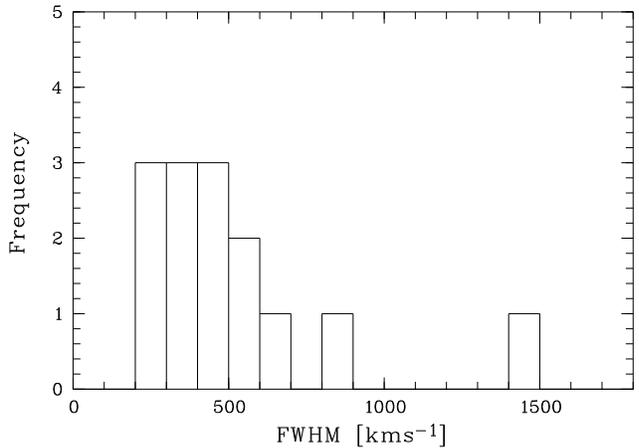}
\caption {Distribution of the line widths ($FWHM$) of the \lya\ emission line of the medium-resolution sample.}
\label{fig_as_fwhm}  
\end{figure}

The full width half maximum (FWHM) values of the \lya\ emission lines are
listed in Table \ref{def_lya2}.  The distribution of the FWHM values corrected
for the instrumental profile is shown in Fig. \ref{fig_as_fwhm}. The observed
values range from 200 \kms\ up to 1500 \kms .  For FDF-4691 and FDF-7539, this
value refers to the envelope of the profile.  Most lines have widths below
$FWHM$ $\approx$ 600 \kms , in agreement with Rhoads et al.
(\cite{rhoads2003}), Dawson et al. (\cite{dawson2004}), and Venemans et al.
(\cite{venemans2004}), who analysed LAEs at a redshift of $z > 5$ and found
line widths of $FWHM < 500$ \kms .

\begin{figure}[htb]
  \resizebox{\hsize}{!}{\includegraphics[clip=true,angle=-90]{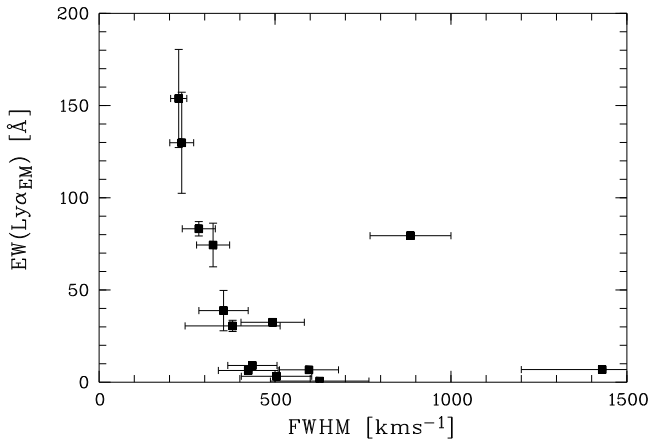}}
\caption {The \lya\ emission line equivalent width as function of the \lya\ emission line width.}
\label{fwhm_ewlya}  
\end{figure}

Figure \ref{fwhm_ewlya} shows the \ewlya\ as a function of the line width of the
emission line. The values of the line width show a possible anti-correlation
with the \lya\ equivalent widths. The outliers with high line width are
FDF-4691 and FDF-7539, both galaxies with double-peak profiles.

\subsection{Slope of the UV continuum}\label{sec_slope}
\begin{figure}[htb]
  \resizebox{\hsize}{!}{\includegraphics[clip=true,angle=-90]{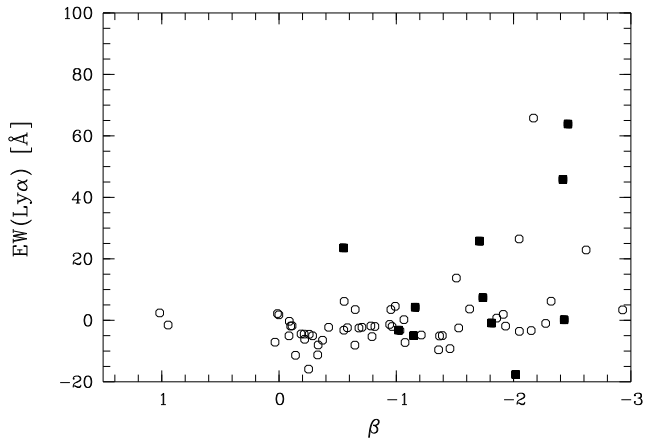}}
\caption {The total \lya\ equivalent width \ewlya\ vs. the continuum slope $\beta$ for the FDF
  spectroscopic sample (circles). Black squares indicate the medium-resolution
  sample.  FDF-4752 is not plotted, since it shows a high equivalent width
  (\ewlya\ = 150 \AA ). FDF-4752 is in the physical vicinity of a QSO
  (FDF-4683). Therefore, its \lya\ emission may be produced differently. }
\label{ew_beta}  
\end{figure}

Studies by, e.g., Calzetti et al. (\cite{calzetti1994}) and Heckman et al.
(\cite{heckman1998}) show that the UV-restframe continuum between 1216 \AA\ 
and 3000 \AA\ of starburst galaxies can be approximated by $F(\lambda) \approx
\lambda ^{\beta}$. Noll et al. (\cite{noll2004}) measured the slope $\beta$
for the galaxies in the FDF spectroscopic survey with 2$<z<$4 in the range
1200 to 1800 \AA , following Leitherer et al. (\cite{leitherer2002}).  The
values of $\beta$ for the medium-resolution sample are listed in Table
\ref{def_lya2}. In Fig. \ref{ew_beta} the continuum slope $\beta$ is plotted
as a function of the total \lya\ equivalent width for the FDF spectroscopic
sample $\beta$ with 2$<z<$4 . Figure \ref{ew_beta} shows that the scatter of the
total \lya\ equivalent width increases towards steeper slopes.  Galaxies with
a continuum slope of $\beta\ >$ -2 have \lya\ equivalent widths below 25 \AA ,
while galaxies with a blue continuum $\beta\ <$ -2 have \lya\ equivalent widths
in the range of \ewlya\ = -20 to 100 \AA .  However, the numbers are too small
to derive firm conclusions.

The measured $\beta$ can be used to derive the dust attenuation for local
starburst galaxies (Heckmann et al. \cite{heckman1998}).  Since the intrinsic
slope depends only weakly on metallicity and starburst age for a
continuous starburst (and an instantaneous starburst within the first 20 Myrs), the
observed slope is determined essentially by the amount of dust (Heckmann et
al.  \cite{heckman1998}; Leitherer et al. \cite{leitherer1999}). However, the
relation of the attenuation and $\beta$ depends on the physical properties of
the ISM and dust in the galaxy.  Therefore, this relationship can be ambiguous
(Noll \& Pierini \cite{noll2005}).

\subsection{Interstellar absorption lines}\label{sec_interabsline}
The equivalent widths of the prominent interstellar SiII $\lambda$1261,
OI/SiII $\lambda$1303, and CII $\lambda$1335 absorption lines were measured
for galaxies in the medium-resolution sample. We restricted ourselves to these
three lines, because the spectral coverage of the medium-resolution spectra
was limited. Only seven galaxies had a continuum SNR that was high enough to
determine the interstellar absorption lines in detail.  The equivalent widths
of the three lines are given in Table \ref{tab_lya_lines5}. The equivalent
widths of the interstellar absorption lines range between -1.1 \AA\ and -3.6
\AA , in good agreement with previous studies (e.g., Noll et al.
\cite{noll2004}).  Moreover, the line widths of the interstellar absorption
lines were measured in the medium-resolution spectra. Only the mean of the
lines SiII $\lambda$1260 and CII $\lambda$1335 are included in Table
\ref{tab_lya_lines5}. SiII/OI $\lambda$1303 was excluded, because the two
lines cannot be separated. The line widths range between 350 and 770 \kms .
These broad line widths are typical for high-redshift galaxies. For their sample of about 800 spectra of
high-redshift galaxies  Shapley et al.
(\cite{shapley2003}) found an average velocity dispersion for the low-ionisation
interstellar absorption lines of $FWHM_{\rm LIS}$ = 560 $\pm$ 150 \kms .

 Velocity
offsets of the interstellar absorption lines with respect to the \lya\ 
emission were derived, using the redshift of the \lya\ emission component as a reference. For these measurements, the OI/SiII $\lambda$1303
blend was again excluded.  The galaxies show a negative velocity offset
between the \lya\ emission line and the interstellar absorption lines (see
Table \ref{tab_lya_lines5}), indicating either a redshifted \lya\ emission or
blue-shifted interstellar absorption lines, or a combination of both.  The
mean velocity offset is $\approx$ -580 \kms , in good agreement with the
values found in other studies (e.g., Shapley et al. \cite{shapley2003}). If we
assume that the redshifts of Noll et al. (\cite{noll2004}) represent the
systemic redshift of the objects, we derive a mean velocity offset of the
interstellar absorption lines of $\approx$ -150 \kms . Therefore, the \lya\ 
emission would be redshifted by $\approx$ 430 \kms , if the Noll et al.
(\cite{noll2004}) measurements indeed represent the systemic redshift.
Adelberger et al. (\cite{adelberger2003}) find for a sample of high-redshift
galaxies that the \lya\ emission is redshifted by 310 \kms\ with respect to
the optical nebular emission lines, while the low-ionisation interstellar
absorption are blueshifted by -150 \kms .

\begin{figure}[htb]

  \resizebox{\hsize}{!}{\includegraphics[clip=true,angle=-90]{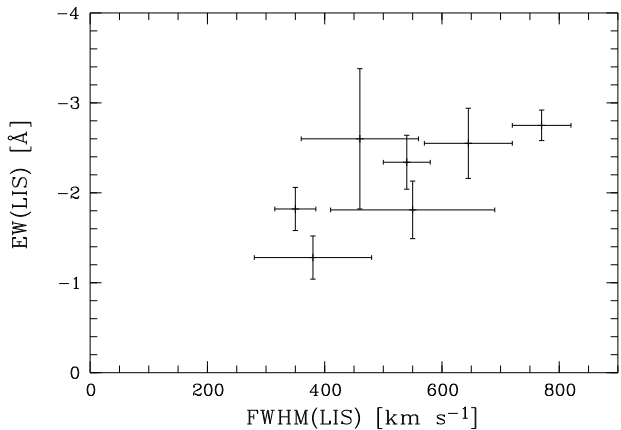}}
\caption {The equivalent width  of the interstellar absorption lines
  plotted as a function of the line width of the interstellar absorption
  lines.}

\label{fwhmlis_ewlis}
\end{figure}

In Fig. \ref{fwhmlis_ewlis} the mean equivalent widths of the interstellar
absorption lines SiII $\lambda$1261 and CII $\lambda$1335 are plotted as a
function of the mean line widths.  A correlation is indicated in Fig.
\ref{fwhmlis_ewlis}.  The strength of the saturated ISM absorption lines does
not strongly depend on the column density, but on the covering fraction and on
the velocity dispersion of the ISM (Shapley et al. \cite{shapley2003}).  The
covering fraction of the low-ionisation interstellar absorption lines can be
measured by the observed residual intensities at the line position, and the
velocity dispersion by the line width.  While the correlation in Fig.
\ref{fwhmlis_ewlis} suggests a significant influence of the gas kinematics,
the scatter indicates that the covering fraction also plays a role. The
galaxies for which the \ewlis\ and \fwhmlis\ could be measured have a lower
average \lya\ equivalent width (average \ewlya\ = -3.2 \AA ) than the full
medium-resolution sample (average \ewlya\ = 37 \AA ).  This is caused by the
fact that the galaxies with lower observed \lya\ equivalent width have a
brighter UV continuum in the FDF spectroscopic survey (see also Table 3 in
Shapley et al. \cite{shapley2003}).  It remains unclear whether the tentative
correlation found above also applies to galaxies with strong \lya\ emission.

\begin{table*}
\caption{ Equivalent widths of the prominent interstellar absorption lines. The
  equivalent widths were measured in the medium-resolution spectra, except
   EW(OI/SiII $\lambda$1303)  and EW(CII $\lambda$1335)  for FDF-5744, which
  were measured in the low-resolution spectra. The
   velocity offset $\delta$v between the \lya\ line  and the prominent low-ionisation 
  interstellar absorption lines and  the widths \fwhmlis\ of the interstellar absorption 
  lines  are also included. The average values for SiII $\lambda$1261 and
  CII $\lambda$1335 are given.}
\vspace{2mm}
\label{tab_lya_lines5}
\centering
\begin{tabular}{c|c|c|c|c|c}
\hline
ID & EW(SiII $\lambda$1261)  & EW(OI/SiII $\lambda$1303)  &
EW(CII $\lambda$1335)  & \fwhmlis\  & $\delta$v
\\
 &  [\AA ]  &  [\AA ] &
 [\AA ] & [\kms ] &  [\kms ]
\\
\hline
1337& - 2.03 $\pm$ 0.42 & - 1.68 $\pm$ 0.4 & - 1.62 $\pm$ 0.25 & 350 $\pm$ 35   & -607 $\pm$ 45   \\
5550& - 2.61 $\pm$ 0.27  & - 2.95 $\pm$ 0.70 &   - 2.07 $\pm$ 0.55    & 540 $\pm$ 40 & -620 $\pm$  85   \\
5744& -2.62 $\pm$ 0.43 & -2.61 $\pm$ 1.1  &   -2.57 $\pm$  1.52  & 460 $\pm$  100  & -560 $\pm$ 120     \\
5903&  -2.54   $\pm$ 0.26  & -3.53  $\pm$ 0.29 &  -2.96  $\pm$ 0.22  & 770 $\pm$  50 &  -760 $\pm$ 80\\
6063& -2.52  $\pm$ 0.59 &  -2.84  $\pm$ 0.65 &  -2.57 $\pm$ 0.66 & 645 $\pm$ 75  &  - \\
7539& -1.4   $\pm$ 0.34 & -1.14  $\pm$ 0.3 &  -1.17  $\pm$ 0.35 &  380 $\pm$ 100 & -80 $\pm$ 30    \\
8304&  -1.81 $\pm$ 0.32  &  -1.83 $\pm$ 0.38 &  -2.53  $\pm$ 0.44 & 550 $\pm$ 140 & -860 $\pm$ 140   \\
\hline
\end{tabular}
\end{table*}

\subsection{Morphology}

\begin{figure*}
  \centering \subfigure[FDF-1337]{\includegraphics[width=3cm]{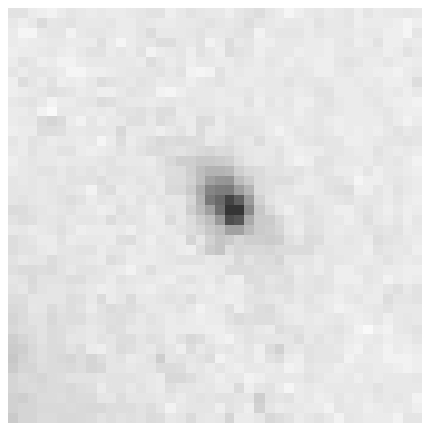}}
  \subfigure[FDF-2384]{\includegraphics[width=3cm]{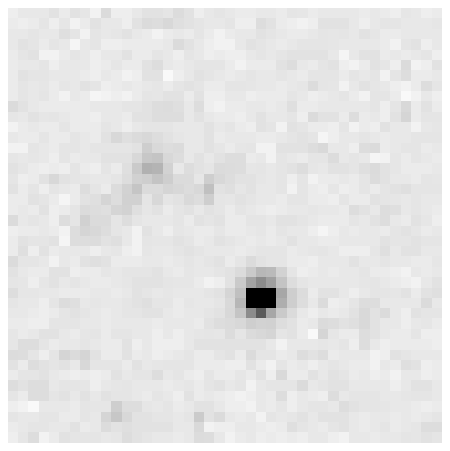}}
  \subfigure[FDF-3389]{\includegraphics[width=3cm]{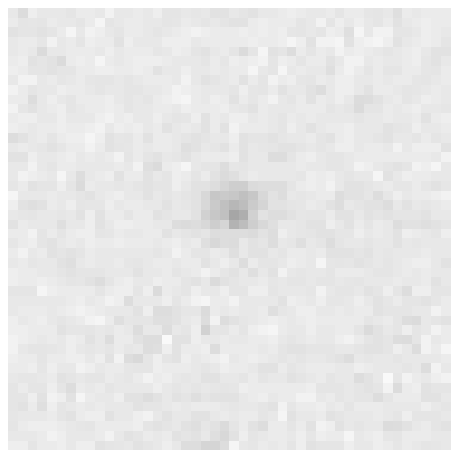}}
  \subfigure[FDF-4691]{\includegraphics[width=3cm]{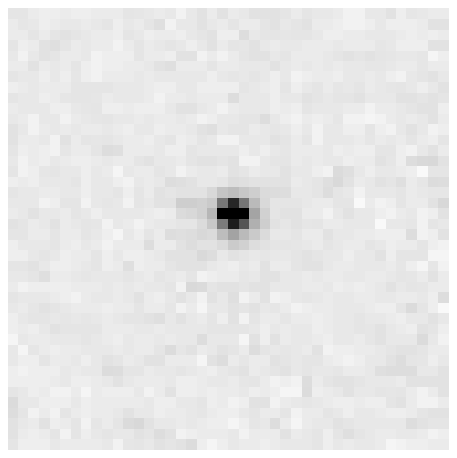}}
  \subfigure[FDF-5215]{\includegraphics[width=3cm]{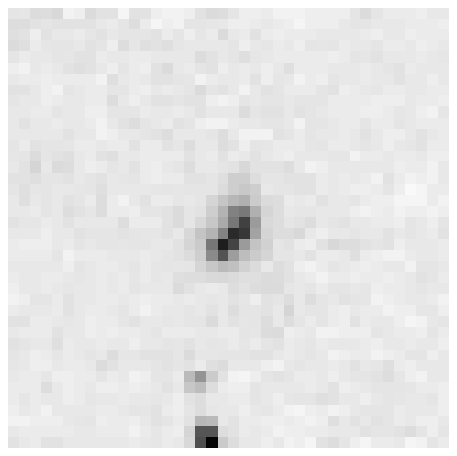}}
  \subfigure[FDF-5550]{\includegraphics[width=3cm]{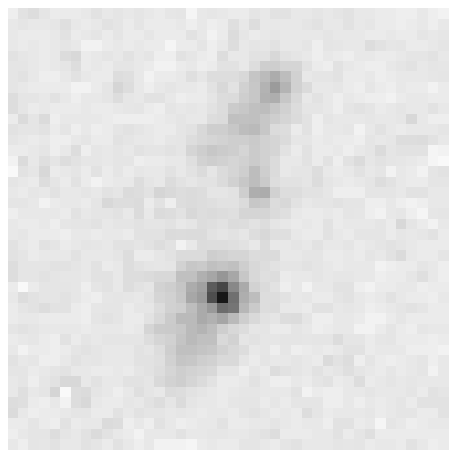}}
  \subfigure[FDF-5744]{\includegraphics[width=3cm]{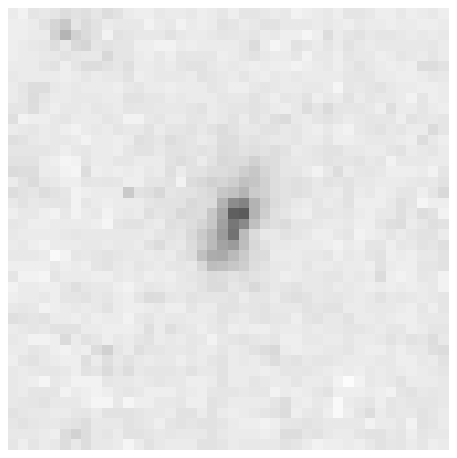}}
  \subfigure[FDF-5812]{\includegraphics[width=3cm]{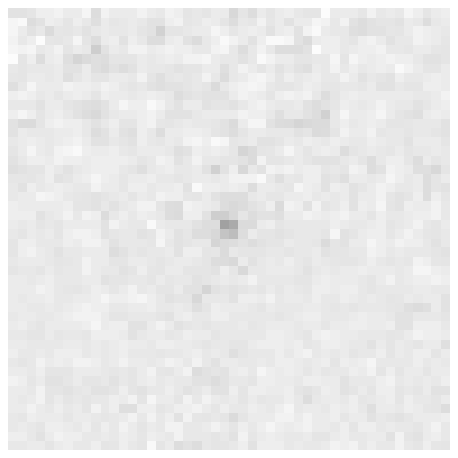}}
  \subfigure[FDF-5903]{\includegraphics[width=3cm]{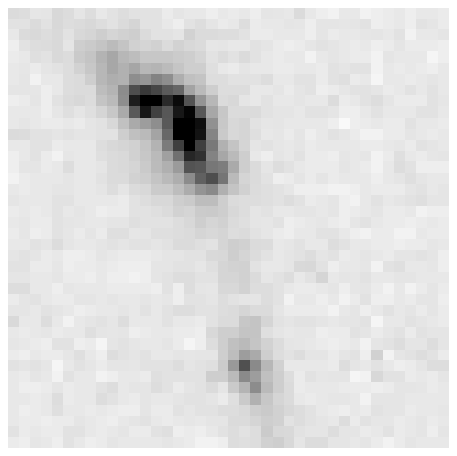}}
  \subfigure[FDF-6063]{\includegraphics[width=3cm]{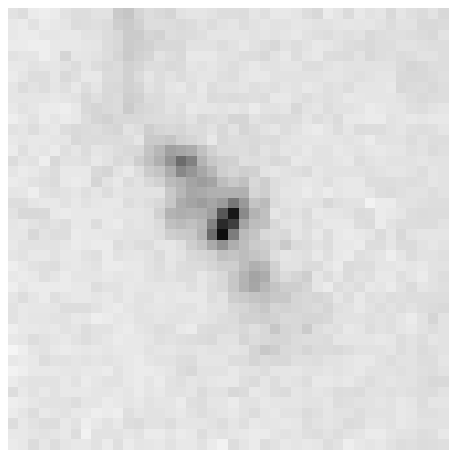}}
  \subfigure[FDF-7539]{\includegraphics[width=3cm]{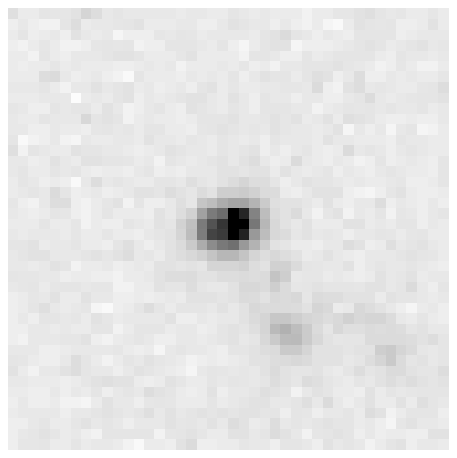}}
  \subfigure[FDF-7683]{\includegraphics[width=3cm]{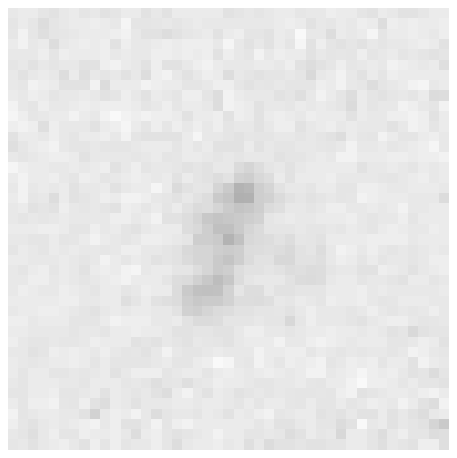}}
  \subfigure[FDF-8304]{\includegraphics[width=3cm]{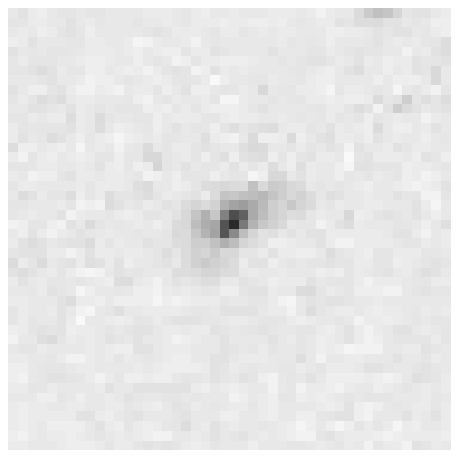}}
\caption {HST/ACS images of the sample of medium-resolution galaxies. Each
  image is 15 $\times$ 15 kpc large, north is up, east is left. }
\label{images_hst}  
\end{figure*}

In Fig. \ref{images_hst} we show HST/ACS F814W images of galaxies of the \lya\ 
medium-resolution sample. Since the HST/ACS image covers only the central part
of the FDF field, not all the galaxies were imaged. The HST/ACS images show
the rest-frame UV of the starburst galaxies, which is dominated by the young
stars. Moreover, as the surface brightness scales with ($z$+1)$^{-4}$, the
extended regions of the galaxies are not easily detected. The observed
diameters of the objects are of the order of $\approx$ 2 kpc, which is typical
of Lyman-break galaxies at a redshift of three (Ferguson et al. 2004).  The
strong \lya\ emitters appear compact (e.g., FDF-4691 or FDF-5215), while the
galaxies with a weak (or no) \lya\ emission line are elongated and irregular
(e.g., FDF-5903 and FDF-6063). We detected neither (strong) spatial extension
of the \lya\ emissions in the two-dimensional spectra nor a spatial offset
between the \lya\ emission and the continuum for all of the members of the
medium-resolution sample, excepting FDF-2384. Figure \ref{images_hst}b shows
that FDF-2384 has two components, a compact one and a diffuse component,
north east of the compact component. The two-dimensional spectrum shows that
the continuum is emitted by the compact object, while the \lya\ emission is
emitted by the diffuse object.

\subsection{Star-formation rates}\label{sec_sfr}
Using the flux of the \lya\ lines, we computed the star-formation rates for the
medium-resolution and low-resolution samples. The line luminosities were
derived by assuming an isotropic emission and Case B recombination, where the
medium was assumed to be optically thick for all Lyman continuum and Lyman
line photons. In this case approximately 2/3 of all recombinations lead to
\lya\ emission (Osterbrock \cite{osterbrock1989}). To derive the
star-formation rates, the calibration of Kennicutt (\cite{kennicutt1998}) was
used. The derived star-formation rates \sfrlya\ for the medium-resolution
sample are given in Table \ref{def_lya2}.

The star-formation rates were also estimated from the UV spectral fluxes,
which have been measured in the low-resolution spectra (if available,
otherwise measured in the medium-resolution spectra). Again, the calibration of
Kennicutt (\cite{kennicutt1998}) was used. The resulting star-formation rates
\sfruv\ range between 1.16 \Msyr and 64 \Msyr (Table \ref{def_lya2}). No
correction for dust absorption was applied to these star-formation rate
values. As shown by Table \ref{def_lya2} the star-formation rates derived from
the \lya\ line are on average lower than the star-formation rates derived from
the UV continuum. For the medium-resolution \lya\ sample, we find a ratio of
\sfrlya\ / \sfruv\ $\approx$ 0.2. Although the conversion of luminosities to
star-formation rates are subject to systematic uncertainties, for most
galaxies it can be concluded that the \lya\ emission is lower than expected
from the star-formation rate derived by the UV continuum. This  agrees
with other measurments of star-formation rates derived from the \lya\ flux and
UV spectral flux (Ajiki et al. \cite{ajiki2003}).

\section{The \lya\ profile}\label{sec_modellya}

\begin{figure*}[htb]
\begin{center}

  \subfigure[FDF-5215]{\includegraphics[width=7.5cm,angle=-90]{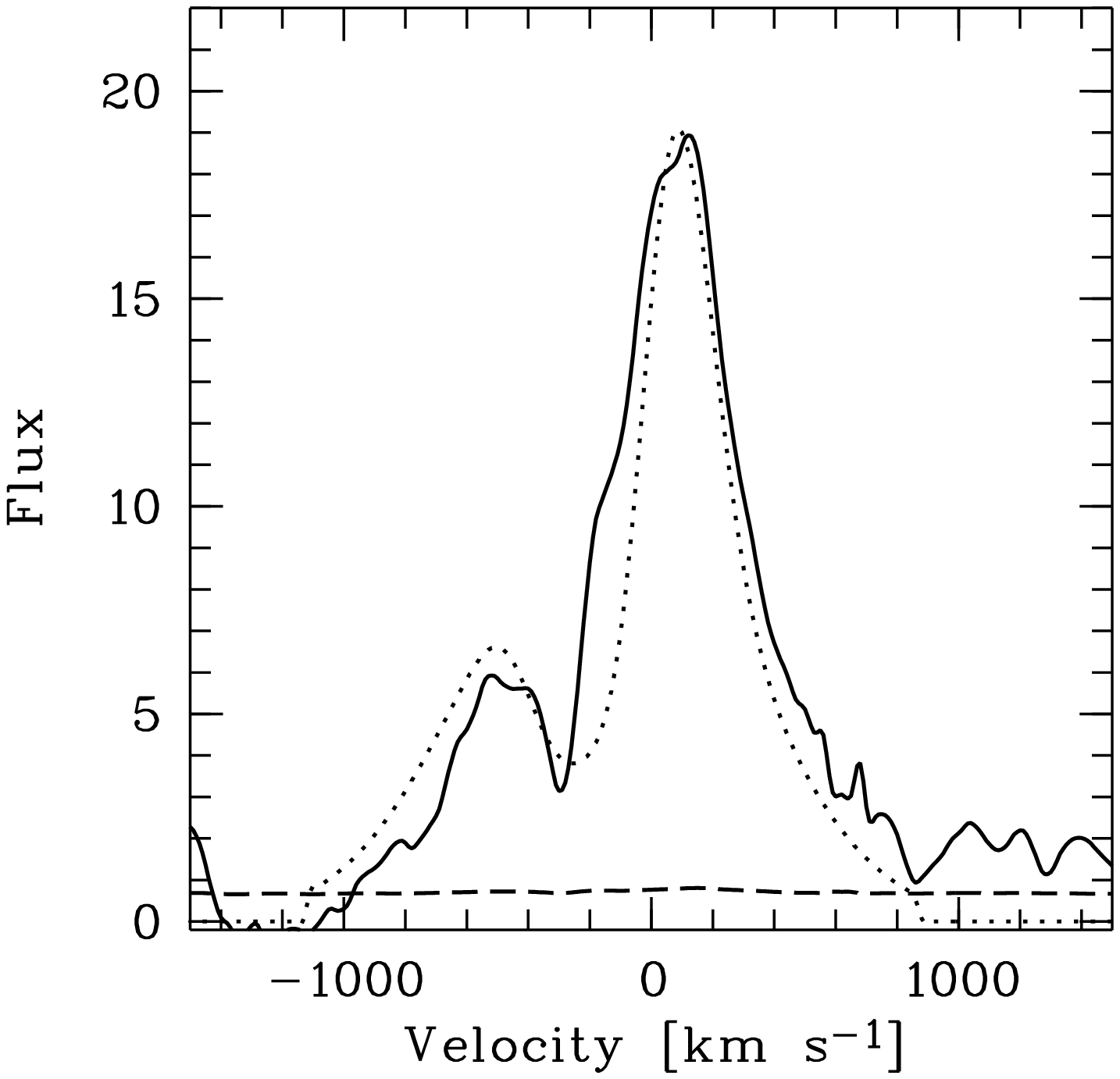}}
  \subfigure[FDF-7539]{\includegraphics[width=7.5cm,angle=-90]{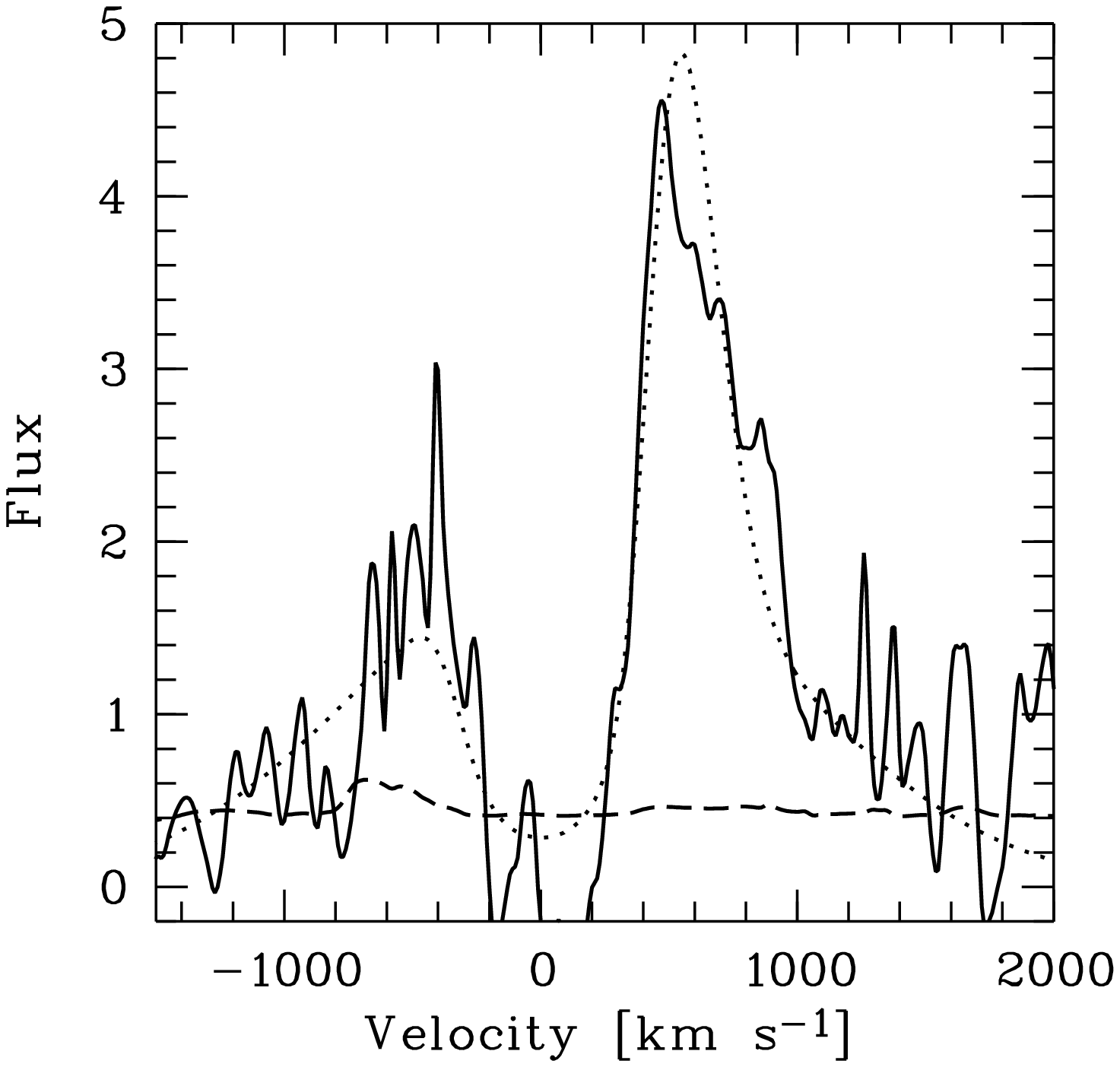}}
\caption {Comparison of the observed \lya\ lines (solid line) of
  FDF-5215 and FDF-7539 and the best-fit theoretical models (dotted line). The
  dashed line indicates the noise level of the observed spectrum.}
\end{center}
\label{FDF5215_lyatheo}  
\end{figure*}

To constrain the physical properties of the Ly$\alpha $ emitting
regions of the galaxies, we compared three of our observed \lya\ profiles with
calculated model profiles using the finite element line formation code of
Richling and Meink\"ohn (Richling \& Meink\"ohn, \cite{richling2001};
Meink\"ohn \& Richling \cite{meinkoehn2002}).  The code is particularly well-suited to calculating the radiative transfer in a non-static scattering
medium. However, the code in its present form is not suitable for high optical
depths.  Therefore, we restricted the finite element modelling to the
double-peaked profiles of FDF-4691, FDF-5215, and FDF-7539. A spherical
two-component model with a central line emission region surrounded by a shell
of neutral HI gas was assumed.  For the emission from the central region, we
assumed a Gaussian emission profile. The model is mainly characterised by the
velocity dispersion of the intrinsic emission line, as well as the velocity
dispersion, the neutral column density, and the outflow velocity of the shell.
The exact geometrical properties of the model are of minor importance for the
computed line profiles (Richling et al. \cite{richling2007}). In Fig.
\ref{FDF5215_lyatheo} the comparison of the \lya\ profiles of FDF-5215 and
FDF-7539 is shown with the theoretical model.  The results for FDF-4691 have
already been described by Tapken et al. (\cite{tapken2004}). The derived fit
parameters are listed for the three galaxies in Table \ref{tab_fem}. While
N$_{\rm{HI}}$ could not be constrained well for FDF-4691, an upper limit can
be found for FDF-5215:  N$_{\rm{HI}}$ cannot exceed N$_{\rm{HI}}$ = $2 \times
10^{16}$ cm$^{-2}$.

\begin{table}
\caption{The derived fit parameters of the finite-element calculations. The velocity dispersion
  of the emission region v$_{\rm{dis}}$(core), the  velocity dispersion of the
shell, the HI column density of the shell  N$_{\rm{HI}}$, and the outflow velocity
of the shell are given.}
\vspace{2mm}
\label{tab_fem}
\centering
\begin{tabular}{c|c|c|c|c}
\hline
ID & v$_{\rm{dis}}$(core) & v$_{\rm{dis}}$(shell) &  N$_{\rm{HI}}$ &
v$_{\rm{outflow}}$  \\
& [\kms ] & [\kms ] & [cm$^{-2}$] &[\kms ] \\
\hline
4691&  600    &   60    &   4 $\times 10^{17}$   &  12 \\
5215& 500 & 125    & $<$ 2 $\times 10^{16}$ & 125 \\
7539 & 1140 & 190 & 2.5 $\times 10^{16}$ & 190 \\ 
\hline
\end{tabular}
\end{table}

FDF-4691, FDF-5215, and FDF-7539 have been modelled with finite element
calculations as well as with Gaussian emission and Voigt absorption
components. Both approaches assume a central emission region producing a
Gaussian emission profile. But the models differ with respect to the radiative
transfer in the absorption component.  The absorption component of the finite
element calculations re-emits the photons. The re-emitted photons are
redistributed in frequency space, leading, e.g, to two emission peaks. The
model with Gauss emission and Voigt absorption assumes that all absorbed
photons are lost. They are either absorbed by dust or  absorbed by
extended neutral clouds, which distribute the photons in physical space and
therefore have surface brightnesses that are too low to be detected (Kunth et al.
\cite{kunth1998}).  The assumption that all absorbed photons are lost may not
be valid for high-redshift galaxies, where the absorbing HI region is compact
with respect to the slit width, and the dust content within the HI region is
expected to be low.  The line fitting of the \lya\ absorption component using
Voigt absorption profiles will lead to an underestimate of the true hydrogen
column density (see also Verhamme et al. \cite{verhamme2006}).

\section{Discussion}\label{sec_discussion}

\subsection{\lya\ profiles}
Meink\"ohn \& Richling (\cite{meinkoehn2002}) and Ahn et al. (\cite{ahn2003})
modelled \lya\ profiles assuming an expanding neutral shell surrounding the
\lya\ emission region. If the shell is static, the profile shows two emission
peaks with the same flux, blueshifted and redshifted with respect to the
systemic redshift. The flux of the blue peak is decreased, if the expansion
velocity is increased.  If the expansion velocity is sufficiently high, the
blueshifted secondary peak disappears. Ahn et al. (\cite{ahn2003}), Ahn
(\cite{ahn2004}), and Verhamme et al. (\cite{verhamme2006}) show, that the
red wing of the \lya\ profile gets more flux. This is caused by \lya\ photons,
which are backscattered from the far side (from an observers point of view) of
the expanding shell, which recedes from an observer. In this case an
asymmetric profile is observed, which can show a secondary emission component
redshifted with respect to the main emission component. Therefore, the model
of an expanding neutral shell surrounding a \lya\ emitting region can
quantitativly reproduce the asymmetric profiles (see also Dawson et al.
\cite{dawson2002}) and the symmetric profiles. A parameter that determines
the morphology of the emission profile is the expansion velocity of the
neutral shell. By a given neutral column density and velocity dispersion of the
neutral shell, a low expansion velocity will lead to a double-peaked \lya\ 
profile, while a higher expansion velocity would lead to an asymmetric
profile. The fact that  asymmetric profiles  in most cases are observed seems to
indicate that the galaxies show an outflow of interstellar HI.  At a redshift
of $z=3$ the mean transmission of the IGM is $T = 0.7$ (Songaila
\cite{songaila2004}). Therefore, we expect that the observed fraction of
double-peak profiles at a redshift of $z = 3$ is not significantly changed by
absorption of the blue part of the profile of the IGM. Most profiles of \lya\ 
emission lines in the literature are asymmetric.  However, double-peaked
profiles are only observable if the signal-to-noise ratio and the spectral
resolution of the spectrum is sufficiently high. A significant number of
spectra with high SNR and resolution have only been observed  for LAEs at $z >
5$, where the absorption of the intergalactic medium is severe ($T<0.2$,
Songaila \cite{songaila2004}). At this redshift a double-peak profile would
not be observed, since the blue peak would be absorped by the IGM. So far,
 double-peaked \lya\ profiles have been observed only at redshift $z \approx 3$
(Fosbury et al. \cite{fosbury2003}; Christensen et al.
\cite{christensen2004}; Venemans et al. \cite{venemans2005}).

\subsection{The origin of the strength of the \lya\ emission in high-redshift galaxies}
We find  no indication of AGN activity in our sample (Sect. 3).  In the
following we assume that the \lya\ emission lines are caused by starburst
activity. In principle, other mechanisms are also possible. However, since the
UV continuum is caused by an ongoing (or recent) starburst, starburst activity
is a reasonable explanation for the observed \lya\ fluxes.  The intrinsic
total \lya\ equivalent width \ewlya\ can be predicted using stellar population
synthesis models if one assumes that the \lya\ photons are produced by
recombination in HII regions ionised by young stars.  The intrinsic total
\lya\ equivalent width \ewlya\ also includes the contribution by stars and
supernovae remnants. For very young ($\approx$ 2 Myr) and very low metallicity
($<$ 4 $\times 10^{-4}$ \Zs ) starburst galaxies, the \lya\ equivalent width
can reach up to 1500 \AA\ (Schaerer \cite{schaerer2003}).  Star-forming
galaxies with moderate age ($>$10 Myr) and metallicity and a initial mass
function (IMF) in the range of
values observed in the local universe have a predicted \ewlya\ in the range
of 50 to 200 \AA\ (Charlot \& Fall \cite{charlot1993}).  However, for
certain star-formation histories, the total \lya\ line can even be observed in
absorption (Charlot \& Fall \cite{charlot1993}). The majority of ionising
photons are produced by early O main-sequence stars, which have a lifetime less
than 4 Myrs.  Stars with lifetimes greater than 10 Myrs do not produce
significant amounts of hydrogen ionising photons. If O stars are absent, the
intrinsic \lya\ flux is very low making the \lya\ equivalent width 
small. While a continuous starburst leads to high ($>$ 100 \AA ) intrinsic
\lya\ equivalent widths, as long as the starburst lasts (Leitherer et al.
\cite{leitherer1999}; Leitherer et al. \cite{leitherer2001}), an exponential decaying star-formation rate (time scale
10 Myrs; Charlot \& Fall \cite{charlot1993}) leads to small equivalent widths
after several 10 Myrs.

Most observed total \lya\ equivalent widths are lower than 20 \AA\ (Sect. 3.1;
see also, e.g, Shapley et al. \cite{shapley2003}; Noll et al.
\cite{noll2004}).  Several effects could lead to a low observed \lya\ 
equivalent width: (a) a suitable epoch of the star-formation history, where
the O stars have already left the main sequence the observed epoch, while B
and A stars producing a strong UV-continuum are still present, (b) orientation
effects resulting in the escape of the \lya\ photons in a different direction,
and (c) absorption of the \lya\ photons by dust and neutral hydrogen (Chen \&
Neufeld \cite{chen1994}). In the following we discuss these three
effects.
  
Shapley et al. (\cite{shapley2003}) and Noll et al. (\cite{noll2004})
present composite spectra of high-redshift galaxies. Both present
composite spectra, which were produced by certain types of high-redshift
galaxies, including galaxies with small observed \lya\ equivalent width.  Strong
stellar wind lines are visible in the composite spectra of galaxies with small
\lya\ equivalent width (Shapley et al. \cite{shapley2003}; Noll et al.
\cite{noll2004}). For example, the CIV \ll\ 1548, 1551 line of the composite
spectrum of the \lya\ subsample group 3 (\ewlya\ = -1 \AA ) of Shapley et al.
(\cite{shapley2003}) displays a P Cygni profile. Only in the observed
UV-restframe spectra of early O main-sequence stars and supergiant OB stars
does the CIV line show a P Cygni profile (Walborn \& Panek \cite{walborn1984};
Walborn \& Nichols-Bohlin \cite{walborn1987}). It is possible that we observe
the galaxies with low \lya\ emission in a short period of their star formation
history (SFH), when the
main sequence O stars, which produce most ionising photons, are no longer
present, and only massive supergiants are present. However, it appears
unlikely that the majority of the high-redshift galaxies with small \lya\ 
equivalent widths are in this phase of their SFH. In addition, Noll et al.
(\cite{noll2004}) find a strong increase of the strength of the \lya\ 
emission from $z \approx 2.3$ to $z \approx 3.2$. There is no reason for the
galaxies at a redshift of $z \approx 3.2$ still to be forming stars, but the
galaxies at $z \approx 2.3$ just turned off their star-formation.  Therefore,
we conclude that an SFH, where the young OB stars are absent but enough B and
A stars produce a strong UV continuum, cannot explain the low observed
\lya\ fluxes easily.  However, one needs multi-wavelength data to derive the
intrinsic \lya\ flux in detail (for a discussion of the determination of the
intrinsic \lya\ flux, see Schaerer \& Pello \cite{schaerer2005}). Therefore, we
cannot rule out that some of the observed variety of the \lya\ emission in
high-redshift galaxies is caused by a different intrinsic \lya\ strength.
 
Another explanation for the observed small \lya\ equivalent widths can be an
orientation effect. In this case the \lya\ photons are produced in huge
numbers and escape the galaxy but are emitted in a direction not
coinciding with the line of sight to us.  As discussed by Charlot \& Fall
(\cite{charlot1993}) and Chen \& Neufeld (\cite{chen1994}), a certain geometric
configuration could lead to an anisotropic emission of the \lya\ photons. The
strength of the \lya\ line depends on the angle at which we observe the
galaxy. Although such orientations effects may play a role for individual
galaxies, the observed correlation between the \lya\ equivalent widths and the
strength of the interstellar low-ionisation lines (discussed below) cannot be
explained if orientation effects are the only reason for the small \lya\
equivalent widths.
  
Since the special stellar population and orientations effects provide no
plausible explanation for the small observed \lya\ equivalent widths, we conclude, that the
most reasonable explanation for the variation in the observed \lya\ equivalent
widths is the combined effect of dust absorption and resonance scattering of
\lya\ photons in neutral hydrogen, enriched with dust (Charlot \& Fall
\cite{charlot1993}; Chen \& Neufeld \cite{chen1994}).  As pointed out by Chen
\& Neufeld (\cite{chen1994}), this mechanism can also explain the strong \lya\ 
absorption seen in many high-redshift galaxies, since the combined effect of
dust absorption and resonance scattering also affects the stellar continuum.
The absorption of the \lya\ photons depends on the total amount of dust, the
neutral gas, the spatial distributation of the dust relative to the gas, and
the kinematical properties. If no neutral gas is present, \lya\ photons and
UV-continuum photons are affected by the same amount of dust.

The analysis of composite spectra of high-redshift galaxies shows the
importance of the ISM for the strength of the \lya\ emission: Shapley et al.
(\cite{shapley2003}) and Noll et al. (\cite{noll2004}) find a strong
anti-correlation between the equivalent width of \lya\ \ewlya\ and the
equivalent width of the low-ionisation metallic interstellar absorption lines
\ewlis . In Lyman-break galaxies the latter are blueshifted with respect to
the system velocity, which is explained by a galaxy-wide outflow, the
superwind (Adelberger et al. \cite{adelberger2003}; Shapley et al.
\cite{shapley2003}).  Observations of Lyman-break galaxies favours a scenario
where clouds of cold neutral gas (which are traced by the low-ionisation
lines) are embedded in hot, ionised gas (Shapley et al. \cite{shapley2003}).
The star-formation activity accelerates the cold gas outwards. The cold
neutral gas phase of the superwind influences the intrinsic \lya\ emission line
strongly (see also  Ferrara \& Ricotti \cite{ferrara2006} for a model for
gas outflow that explains the observed properties of Lyman-break galaxies, including the
strengths of their \lya\ emission).  The properties of the superwind will  
determine the observed \lya\ flux. One
can describe this superwind by the following model parameters: the velocity
dispersion $b$, the neutral column density of the hydrogen \nhi , the mean
outflow velocity $v$, and the spatial covering fraction. Concerning the
radiative transfer, the superwind is rather similar to the expanding shell
model used by us in Sect. 5.1 to simulate the formation of the observed \lya\ 
profiles. Therefore, the agreement of the observed and computed profiles can
also be regarded as support of the superwind model. In principle, the observed
properties of the superwind (like the mean outflow velocity) can be used to
constrain the parameters of the expanding shell model, which decribes the \lya\ 
profiles (Verhamme et al. \cite{verhamme2006}).  This will hopefully help for
understanding  the observed properties of the \lya\ emission in more detail.

\section{Conclusions}\label{sec_conclusion}
We analysed a sample of 16 restframe UV-continuum selected 2.7 $ < z < $ 5
galaxies using medium-resolution FORS2 spectra.  The \lya\ lines range from pure absorption (\ewlya\ =
-17 \AA ) to strong emission (\ewlya\ = 153 \AA). Most of the \lya\ lines show
an asymmetric profile, with a sharp drop on the blue side and an extended red
wing.  Three \lya\ lines display a double-peaked profile with two separate
emission lines.  The profiles were compared to calculated  profiles computed
by a radiative transfer code.  Both types of profiles, the asymmetric and the
double-peaked profiles, can be explained by a uniform model of an expanding
shell of neutral hydrogen surrounding a compact starburst region (see also Ahn
et al. \cite{ahn2001}, \cite{ahn2003}).  One parameter, which determines the
morphology of the profile, is the expansion velocity of the neutral shell.  A
low expansion velocity leads to a double-peaked \lya\ profile, while a higher
expansion velocity would lead to an asymmetric profile. The \lya\ emission
strengths of our target galaxies are found to be determined by the amount of
dust and the kinematics of the outflowing material.  Broad, blueshifted,
low-ionisation interstellar absorption lines were detected. They indicate a
galaxy-scale outflow of the ISM. The strengths of these ISM lines were found to
be partly determined by the velocity dispersion of the outflowing medium.

\begin{acknowledgements}
  Our research has been supported by the German Science Foundation DFG (SFB
  439). We thank Daniel Schaerer, Anne Verhamme, and Klaus Meisenheimer for
  valuable discussions. We would also like to thank the referee for his/her
  insightful comments that improved the paper. We thank the Paranal staff for
  having carried out the service mode observations and  Maurilio Pannella for
  providing the HST images.
\end{acknowledgements}

\end{document}